%% file: main.tex
\documentclass[sigconf,balance=false,authorversion,nonacm]{acmart}

\AtBeginDocument{%
  \providecommand\BibTeX{{%
    Bib\TeX}}}

\input{preamble}

\begin{document}
\sloppy
\title{Differentially Private Speaker Anonymization}


\author{Ali Shahin Shamsabadi}\authornote{Most of this work was done during the internship of Ali Shahin Shamsabadi at Inria / University of Lille.}
\affiliation{%
  \institution{The Alan Turing Institute \\Vector Institute}
  \country{}
}
\email{a.shahinshamsabadi@turing.ac.uk}

\author{Brij Mohan Lal Srivastava}
\affiliation{%
  \institution{Université de Lille, Inria, CNRS, Centrale Lille, UMR 9189 - CRIStAL, F-59000 Lille, France}
  \country{}
}

\author{Aurélien Bellet}
\affiliation{%
  \institution{Université de Lille, Inria, CNRS, Centrale Lille, UMR 9189 - CRIStAL, F-59000 Lille, France}
  \country{}
}

\author{Nathalie Vauquier}
\affiliation{%
  \institution{Université de Lille, Inria, CNRS, Centrale Lille, UMR 9189 - CRIStAL, F-59000 Lille, France}
  \country{}
}
\author{Emmanuel Vincent}
\affiliation{%
  \institution{Université de Lorraine, CNRS, Inria, LORIA, F-54000 Nancy, France}
  \country{}
}

\author{Mohamed Maouche}
\affiliation{%
  \institution{Université de Lille, Inria, CNRS, Centrale Lille, UMR 9189 - CRIStAL, F-59000 Lille, France}
  \country{}
}

\author{Marc Tommasi}
\affiliation{%
\institution{Université de Lille, Inria, CNRS, Centrale Lille, UMR 9189 - CRIStAL, F-59000 Lille, France}
  \country{}
}

\author{Nicolas Papernot}
\affiliation{%
  \institution{Vector Institute \\ University of Toronto}
  \country{Canada}
}

\renewcommand{\shortauthors}{Shahin Shamsabadi et al.}

\input{arxiv}

\clearpage

\appendix
\input{appendix}

\end{document}

%% file: preamble.tex
\usepackage{algorithmic}
\usepackage{graphicx}
\usepackage{textcomp}
\usepackage{booktabs}
\usepackage{multirow}
\usepackage{xcolor}
\usepackage{xspace}
\usepackage{hyperref}
\usepackage{pgfplotstable}
\usepackage{url}                
\usepackage{xcolor}             
\usepackage[]{hyperref}         
\usepackage[T1]{fontenc}
\usepackage[utf8]{inputenc}
\usepackage{tikz,pgfplots}

\usepgfplotslibrary{statistics}
\pgfplotsset{compat=1.8}
\hypersetup{                    
  colorlinks,
  linkcolor={green!80!black},
  citecolor={red!70!black},
  urlcolor={blue!70!black}
}

\def\BibTeX{{\rm B\kern-.05em{\sc i\kern-.025em b}\kern-.08em
    T\kern-.1667em\lower.7ex\hbox{E}\kern-.125emX}}

\definecolor{DP}{RGB}{175,0,3}
\definecolor{Org}{RGB}{233,107,34}

\newtheorem{definition}{Definition}
\DeclareMathOperator*{\range}{range}

\newcommand{\myparagraph}[1]{\vspace{1ex}\noindent{\bf #1}}

\newcommand{\map}{\mathcal{A}}
\newcommand{\enc}{\mathcal{E}}
\newcommand{\noise}{\mathcal{N}}
\newcommand{\dec}{\mathcal{D}}
\newcommand{\B}{\mathcal{B}}
\newcommand{\M}{\mathcal{M}}
\newcommand{\T}{\mathcal{T}}
\newcommand{\p}{\mathbf{p}}
\newcommand{\bnm}{\mathbf{B}}
\newcommand{\bn}{\mathbf{b}}
\newcommand{\h}{\mathbf{h}}
\newcommand{\x}{\mathbf{x}}

\newcommand{\DP}{{\texttt{Anon+DP}}\xspace}
\newcommand{\Anon}{{\texttt{Anon}}\xspace}
\newcommand{\AnonPC}{{\texttt{Anon+PC}}\xspace}
\newcommand{\AnonDPBN}{{\texttt{Anon+DP\_BN}}\xspace}
\newcommand{\AnonDPPitch}{{\texttt{Anon+DP\_Pitch}}\xspace}

\def\checkmark{\tikz\fill[scale=0.4](0,.35) -- (.25,0) -- (1,.7) -- (.25,.15) -- cycle;} 
\DeclareMathOperator*{\argmax}{argmax}

%% file: arxiv.tex
\begin{abstract}
Sharing real-world speech utterances is key to the training and deployment of voice-based services. However, it also raises privacy risks as speech contains a wealth of personal data. 
Speaker anonymization aims to remove speaker information from a speech utterance while leaving its linguistic and prosodic attributes intact. State-of-the-art techniques operate by disentangling the speaker information (represented via a speaker embedding) from these attributes and re-synthesizing speech based on the speaker embedding of another speaker. Prior research in the privacy community has shown that anonymization often provides brittle privacy protection, even less so any provable guarantee. In this work, we show that disentanglement is indeed not perfect: linguistic and prosodic attributes still contain speaker information. 
We remove speaker information from these attributes by introducing differentially private feature extractors based on an autoencoder and an automatic speech recognizer, respectively, trained using noise layers.
We plug these extractors in the state-of-the-art anonymization pipeline and generate, for the first time, private speech utterances with a provable upper bound on the speaker information they contain. We evaluate empirically the privacy and utility resulting from our differentially private speaker anonymization approach on the LibriSpeech data set. Experimental results show that the generated utterances retain very high utility for automatic speech recognition training and inference, while being much better protected against strong adversaries who leverage the full knowledge of the anonymization process to try to infer the speaker identity.

\end{abstract}

\maketitle

\section{Introduction}

Recent advances in automatic speech recognition (ASR) \cite{graves2013speech,amodei2016deep} have enabled the deployment of voice-based services such as dictation, voice search and voice assistants in our daily life \cite{clark2019state}. In these applications, speech is collected by service providers and third-party contractors\footnote{See, e.g., \url{https://www.bbc.com/news/technology-31296188}.} to process user queries (\emph{inference}), but also before deployment to train ASR systems on real, diverse, annotated speech data (\emph{training}) and thereby achieve state-of-the-art performance \cite{tuske2014data,kawakami2020learning}.
However, speech data is very sensitive, not only through the linguistic content (what is being said) but first and foremost because it is a biometric modality that can identify the speaker \cite{jain2000biometric}.
Automatic speaker identification (ASI) and speaker verification (ASV) techniques can identify and distinguish speakers in large populations with low error \cite{bimbot2004tutorial,7298570}.
Malicious parties with access to the speech data of a victim
can impersonate him/her or assemble fake recordings that he/she never said \cite{brasser2018voiceguard}.
Disseminating speech data thus entails significant privacy and security risks. The last two years have seen the rise of privacy issues in the agenda of the speech processing community, as evidenced by the creation of a special interest group of the International Speech Communication Association,\footnote{\url{https://www.spsc-sig.org}} the launch of the VoicePrivacy initiative \cite{tomashenko2020introducing}, and ongoing efforts to understand the requirements of effective privacy preservation for speech data \cite{nautsch2019preserving} in light of recent regulation \cite{nautsch2019gdpr}.

In this work, we are interested in the problem of \emph{speaker anonymization}, which aims to conceal the speaker's identity (privacy) while preserving the linguistic and prosodic (intonation, stress and rhythm) content as well as the diversity of speech (utility). This problem was the focus of the recent VoicePrivacy challenge \cite{vpc2022csl}. A successful speaker anonymization approach enables users to freely share their speech data with service providers for both inference and training purposes, while concealing their identity.
It is important to note that, as in the VoicePrivacy challenge, we seek to preserve all linguistic content and do not address the problem of protecting the personally identifiable information that it may contain (e.g., names, addresses, credit card numbers) --- this could be done using privacy-preserving ASR methods \cite{ahmed2020preech}, for instance.
Speaker anonymization can benefit many real-world applications beyond ASR-based services.  For instance, in France, legal cases can be broadcasted on TV provided that the speakers’ voices are transformed so as to prevent re-identification while preserving the linguistic and prosodic content. Call centers which record customer calls are facing the same problem.

Speaker anonymization cannot be addressed by cryptographic solutions as they provide confidentiality (i.e., only the data owner can observe data) \cite{zhang2019encrypted}, not privacy (i.e., protecting what can be inferred about the speaker identity from speech data). For example, cryptographic solutions make human annotation impossible \cite{tomashenko2020introducing}.
A naive approach to speaker anonymization consists in transcribing the speech into text using an ASR model followed by a text-to-speech (TTS) system that re-synthesizes speech from the text transcription. While this ASR+TTS approach perfectly conceals the speaker identity, it destroys the utility of speech in several ways: in addition to the incorrect linguistic content induced by (unavoidable) ASR errors, the original prosodic attributes (intonation, stress, and rhythm) are lost, and the variability of synthesized speech output by TTS is very limited, especially when the TTS system can only generate a few voices.
Due to this limited diversity, ASR models trained on synthetic speech generated by TTS perform poorly when applied to real speech \cite{li2018training,chen2020improving}.

\looseness=-1 By contrast, state-of-the-art speaker anonymization methods seek to separate the speaker identity information from the linguistic and prosodic content so as to generate speech where only the identity information has been removed \cite{fangspeaker,tomashenko2020introducing,srivastava2020design,srivastava:hal-03197376}. These methods rely on the extraction of three types of features from a speech recording: (i) a speaker embedding (typically an x-vector \cite{snyder2018x} extracted from an intermediate layer of an ASI model) which encodes the characteristics of the speaker's voice, (ii) a sequence of bottleneck (BN) features \cite{yu2011improved} (a low-dimensional phonetic representation extracted from an intermediate layer of an ASR model) that captures fine-grained linguistic information (as opposed to the possibly erroneous word sequence in the ASR+TTS approach), and (iii) a sequence of pitch features (i.e., the signal's fundamental frequency) which conveys prosodic information \cite{gussenhoven2004phonology}. Speaker anonymization is then realized by re-synthesizing speech from the BN and pitch features of the original speech recording and a replaced speaker embedding corresponding to another (real or pseudo) speaker.
While this general approach has been quite successful and achieves good practical performance \cite{srivastava:hal-03197376}, there remains a lot of room for improvement in protecting against concrete attacks \cite{slicing}. In particular, the disentanglement of speaker information is not perfect: linguistic and prosodic features are known to contain residual identity information \cite{fangspeaker} which can propagate to the anonymized speech and be used by an adversary to re-identify speakers (see Section~\ref{sec:val}). Furthermore, the effectiveness of anonymization is evaluated only empirically: even if the evaluation is performed using state-of-the-art ASI or ASV techniques and takes into account some auxiliary information that the adversary may have \cite{Srivastava2020a}, there is no guarantee that the resulting speech cannot be de-anonymized using better attacks.

\looseness=-1 In this paper, we propose to use ideas from $\varepsilon$-differential privacy ($\varepsilon$-DP) \cite{DP2006}, a rigorous mathematical framework to quantify the information leakage of algorithms, to design more robust speaker anonymization techniques. Following the pipeline of state-of-the-art methods described above, we introduce differentially private pitch and BN feature extractors that can bound the risk of the speaker identity leaking through the prosodic and linguistic attributes used to re-synthetize speech. While it is easy to enforce DP by adding random noise directly to the original features, this naive approach destroys the linguistic and prosodic content that we wish to preserve. Instead, we carefully design machine learning-based pitch and BN feature extractors, and train them to retain the desired information while adding the necessary amount of noise to get DP guarantees via a Laplace noise layer.
Specifically, for pitch, we introduce a novel autoencoder (an encoder-decoder network) that learns
to reconstruct the input pitch at the output by optimizing
an original reconstruction loss function
designed to preserve the global pitch dynamics which conveys prosodic information (e.g., pitch increases when asking a question) while the noise perturbs the local variations that are more speaker-specific \cite{adami2003moeling,peskin2003using,dehak2007modeling,mary2008extraction}.
For BN features, we train a deep ASR acoustic model
to learn features that retain as much as possible the phonetic information needed to decode the linguistic content, while the noise helps to remove the residual speaker information.
Regarding the speaker embedding, we simply choose a public x-vector (provided by the VoicePrivacy challenge \cite{srivastava:hal-03197376}) randomly and independently of the input utterance so that it does not contain any information about the original speaker, following \cite{srivastava2020design}.
Plugging our private feature extractors into the full speaker anonymization pipeline,
we obtain a differentially private version of the above state-of-the-art speaker anonymization approach.

\looseness=-1 Our approach satisfies a rigorous analytical DP guarantee that upper bounds the leakage of the speaker identity by the parameter $\varepsilon$ (the smaller $\varepsilon$, the stronger the privacy guarantee). Being a worst-case measure, DP however gives a conservative privacy guarantee: there is often a large gap with what adversaries can infer in practical settings \cite{DP_eval,jagielski2020auditing,nasr2021adversary}. To complement our analytical guarantees, we lower bound the leakage of the speaker identity by empirically evaluating the success of concrete adversaries designed to be as close to the worst case as possible given realistic knowledge. We conduct a two-step evaluation of empirical privacy and utility.
First, we evaluate the ability of an adversary to re-identify a known speaker from pitch and BN features by training an ASI model directly on these features. Our results show that the features output by our proposed DP extractors provide much better protection than the standard features used in previous work.
Second, plugging our feature extractors into the state-of-the-art speaker anonymization technique \cite{tomashenko2020introducing,srivastava2020design}, we show that we can generate speech utterances which empirically preserve better the privacy of speakers (even when using rather large values for $\varepsilon$) at only a small cost in utility.
Here, utility is measured by the word error rate (WER) of an ASR system trained and tested on anonymized speech, and privacy by the equal error rate (EER) of an adversary that uses a state-of-the-art ASV system trained on a large corpus of anonymized utterances.
Low WER and high EER indicate that the speech generated with our approach can be shared, stored, annotated and used to train ASR models for voice-based services, while protecting the speaker identity.

In summary, our contributions advance the state-of-the-art in speaker anonymization as follows: \begin{itemize}
    \item \looseness=-1 We empirically demonstrate that the BN and pitch features used by current speaker anonymization methods contain a lot of speaker information by mounting an attack in which we train ASI models directly on these features. Our attack achieves 97\% and 37\% accuracy on BN and pitch features, respectively, on the 921 speakers from the LibriSpeech data set.
    \item \looseness=-1 We introduce DP pitch and BN feature extractors that remove the speaker identity while preserving the linguistic and prosodic information. We show that our extractors provide analytical $\varepsilon$-DP guarantees.
In addition to this, our DP extractors with $\varepsilon=1$ reduce the accuracy of the above attack on BN and pitch to 14\% and 5\%, respectively. We show that our BN features can be shared instead of raw utterances to perform ASR training and inference with negligible effect on the WER. On LibriSpeech, our DP BN extractor with $\varepsilon=1$ achieves $6\%$ WER, compared to $5\%$ with the original BN features. 
    \item Finally, we synthesize speech from our DP extractors and compare against state-of-the-art speaker anonymization. We evaluate the empirical privacy by mounting an attack in which we train ASV models on anonymized utterances.
Our results demonstrate that the state-of-the-art speaker anonymization pipeline still leaks speaker identity information. Remarkably, in addition to its analytical privacy guarantees, our approach provides much better empirical privacy while utility remains very high. On LibriSpeech, our DP speaker anonymization scheme achieves an EER and a WER of 30\% and 7\%, respectively, whereas the state-of-the-art scheme achieves an EER and a WER of respectively 15\% and 5\%. 
\end{itemize}

\section{Background}
\label{sec:Background}

In this section, we introduce background concepts in speech processing and differential privacy.

\subsection{Speech Processing}
\label{sec:background_speech}
Speech data contain speaker information as well as linguistic and prosodic attributes.
State-of-the-art speaker anonymization relies on automatic speaker recognition, automatic speech recognition and pitch estimation for extracting speaker information, linguistic and prosodic attributes, respectively. 
We give an overview of these techniques, which will be used both in the design of our approach and in our empirical evaluation.

\myparagraph{Automatic speaker recognition} is the task of recognizing the speaker of a given speech utterance.  Existing techniques can be categorized into automatic speaker verification (ASV), i.e., authenticating the identity claimed by the speaker, and automatic speaker identification (ASI), i.e., determining the identity within a set of known speakers \cite{bimbot2004}.
ASI relies on training a neural network based speaker classifier on utterances from multiple speakers and later using that network to classify the identity of each test utterance as one of the known training identities.
As opposed to the closed-set ASI task, ASV is an open-set (rejection/acceptance) task which comprises two successive phases: enrollment and authentication.
In the former, a speaker embedding is extracted from one or more enrollment utterances spoken by the speaker whose identity is being claimed. The most popular embeddings called x-vectors \cite{snyder2018x} are obtained from an intermediate layer of a neural network trained to perform ASI. In the latter phase, the x-vector extracted from the utterance of an unknown speaker (called trial utterance) is compared with the x-vector of the speaker whose identity is being claimed. 
and a log-likelihood ratio score is computed by probabilistic linear discriminant analysis (PLDA) \cite{kenny2010}. The ASV system then decides whether the trial utterance is from that speaker or not by comparing the obtained score with a threshold.

\looseness=-1 \myparagraph{Automatic speech recognition (ASR)}
aims to convert an utterance into its textual content, also called transcription. We describe here the classical monolingual ASR architecture based on separate acoustic and language models, which was used both to extract BN features and to quantify utility in the VoicePrivacy challenge \cite{vpc2022csl}. End-to-end neural ASR architectures and/or multilingual architectures could also be used without loss of generality \cite{champion:hal-02995855}.
The input to ASR is a sequence $\mathbf{O}=[\mathbf{o}_1,\dots,\mathbf{o}_K]^\top \in \mathbb{R}^{K\times A}$ of length $K$ of acoustic feature vectors $\mathbf{o}_k \in \mathbb{R}^A$ derived from speech, e.g., Mel-frequency cepstral coefficients (MFCCs), and the output is the estimated word sequence $\hat{W}$. This problem can be formulated as \cite{yu2016automatic}
\begin{equation}
    \textstyle\hat{W} = \argmax_W P(W | \mathbf{O}).
\end{equation}
In practice, it is infeasible to directly model the conditional distribution of the true word sequence given the acoustic features. Hence, using Bayes' rule, some independence assumptions and the fact that $\mathbf{O}$ is fixed, the ASR problem is reformulated as \cite{manohar2019semi}
\begin{align}
    \hat{W} &= \argmax_W P(\mathbf{O} | W) P(W)/ P(\mathbf{O}) \nonumber \\
    \hat{W} &= \argmax_W P(\mathbf{O} | W) P(W) \nonumber \\
            &= \argmax_W \sum_{S,N} P(\mathbf{O}|S) P(S|N) P(N|W) P(W).
\end{align}
\looseness=-1 Here $P(W)$ is the so-called language model that represents the prior distribution of word sequences, $P(N|W)$ is the lexicon which maps words to the corresponding phoneme sequences, $P(S|N)$ maps a phoneme sequence to the corresponding triphone (i.e., tied context-dependent phoneme) sequence $S=[S_1,\dots,S_K]$, and $P(\mathbf{O}|S)\propto\prod_{k=1}^K P(S_k|\mathbf{O})/P(S_k)$ where the triphone posterior probabilities $P(S_k|\mathbf{O})$ are given by the so-called acoustic model and $P(S_k)$ is the prior probability of each triphone. These models are trained independently and composed together as a graph using finite state transducers. 

\myparagraph{Bottleneck features.} An acoustic model trained for ASR can also be used for other tasks which rely on the phonetic content but do not require a word-level transcription, such as language identification \cite{duroselle:hal-03264085} or keyword spotting \cite{VANDERWESTHUIZEN2022101275}. In such cases, instead of using the acoustic model output (triphone posterior probabilities), a sequence of phonetic features called bottleneck (BN) features is extracted from an intermediate layer of the acoustic model \cite{yu2011improved} and used, possibly in combination with other features, as input to these tasks.

\myparagraph{Pitch estimation.} The fundamental frequency (called pitch and denoted as F0) is the frequency of oscillation of the vocal folds. The vocal folds are the flap-like organ at the upper end of the trachea which controls the air stream emerging from the lungs. The range of pitch is determined by the physiological factors of the vocal folds, such as their mass and length, hence it depends on the speaker and is typically lower for male than female \cite{biemans1998effect}. The pitch sequence governs the {\it intonation} of the spoken utterance. It is a key component of prosody (together with stress and rhythm) which determines the utterance expressiveness.
It is important to note that pitch is the rate of vibration of vocal folds hence it is only defined for voiced phonemes such as /a/, /b/, /z/, etc.
It is pointless to compute pitch for silence, noise or unvoiced regions of an utterance since there is no vibration of the vocal folds, hence it is conventionally zero at these locations.
Pitch estimation is a difficult task due to erroneous observation of harmonics causing pitch doubling/halving \cite{zahorian2008spectral}. It is also difficult to estimate the pitch when the quality of speech is distorted due to noise or channel effects.
We use a fairly robust and widely used algorithm for pitch tracking called YAAPT \cite{kasi2002yet}.

\subsection{Differential Privacy}

Differential Privacy (DP) \cite{DP2006} provides a rigorous probabilistic way to quantify the privacy leakage of an information release process. DP also comes with strong mathematical properties and a powerful algorithmic framework \cite{Dwork2014a}. For these reasons, DP and its variants have become the gold standard notion of privacy in machine learning and many other scientific fields. DP has also seen recent real-world deployments, notably by the US Census \cite{USCensus}.

\looseness=-1 Two main trust models have been considered in DP. The central model assumes the presence of a trusted curator which collects raw data from data owners. In the local model \cite{Kasiviswanathan2008,duchi2013local}, each data owner obfuscates its data locally before sharing it. As we aim to design methods for speakers to anonymize their speech, we place ourselves in the local model. Formally, $\varepsilon$-differential privacy is defined as follows.
\begin{definition}[Local Differential Privacy~\cite{Kasiviswanathan2008,duchi2013local}]
Let $\mathcal{A}$ be a randomized algorithm taking as input a data point in some space $\mathcal{X}$, and let $\varepsilon>0$. We say that $\mathcal{A}$ is $\varepsilon$-differentially private ($\varepsilon$-DP) if for any $\x,\x'\in\mathcal{X}$ and any $S\subseteq\range(\mathcal{A})$:
$$\Pr[\mathcal{A}(\x)\in S] \leq e^\varepsilon\Pr[\mathcal{A}(\x')\in S],$$
where the probabilities are taken over the randomness of $\mathcal{A}$.
\end{definition}

DP essentially requires that the probability of any output does not vary ``too much'' (as captured by $\varepsilon$) when changing the input. The smaller $\varepsilon$, the stronger the privacy guarantee.
In our setting, a data point $\x$ will correspond to a speech utterance and $\mathcal{A}$ will be a speaker anonymization procedure that produces an anonymzed utterance as output. DP then ensures that any two arbitrary utterances will be indistiguinshable (to some extent), and therefore bounds the risk that an adversary observing the output can predict who spoke it.

\looseness=-1 DP possesses a number of desirable properties that we will use in our work. First, any function of an $\varepsilon$-DP algorithm remains $\varepsilon$-DP (\emph{robustness to post-processing}). Second, one can easily keep track of the privacy guarantees across multiple analyses (\emph{composition}). In particular, given $K$ algorithms that satisfy $\varepsilon$-DP, executing them on the same data and releasing their combined outputs is $K\varepsilon$-differentially private.

A standard way to design differentially private algorithms is based on output perturbation. In this work, we will rely on the so-called Laplace mechanism, which consists in adding Laplace noise calibrated to the $\ell_1$-sensitivity of the (non-private) function one would like to compute on the data.

\begin{definition}[Laplace mechanism]
\label{defn:laplace-mech}
Let $f:\mathcal{X}\rightarrow\mathbb{R}^d$ and let the $\ell_1$-sensitivity of $f$ be defined as $$\textstyle\Delta_1(f)=\max_{\x,\x'\in\mathcal{X}}\|f(\x)-f(\x')\|_1.$$
Let $\eta=[\eta_1,\dots,\eta_d]\in\mathbb{R}^d$ be a vector where each ${\eta_i\sim \text{Lap}(\Delta_1(f)/\varepsilon)}$ is drawn from the centered Laplace distribution with scale $\Delta_1(f)/\varepsilon$. Then, $\mathcal{A}(\cdot)=f(\cdot)+\eta$ is $\varepsilon$-DP.
\end{definition}

In our work, we will use the Laplace mechanism to construct a differentially private transformation of speech utterances for speaker anonymization.
It is important to note that DP enforces a stronger notion of privacy than what we aim to achieve: as explained above, it entails hiding the speaker identity but may also suppress other information that we wish to preserve. In particular, adding Laplace noise to raw speech destroys linguistic and prosodic information. Instead, we will show that applying DP at an intermediate layer of carefully designed feature extractors trained to preserve linguistic and prosodic information, we can successfully conceal the speaker identity while retaining the usefulness of anonymized utterances for ASR training and inference. This is in line with other recent work on using DP to hide specific attributes from image \cite{image_DP} and text data \cite{Beigi_write,DP_text,DP_WE}.

\section{Problem Statement and Threat Model}
\label{sec:ProblemFormulation}
Inspired by the VoicePrivacy challenge \cite{tomashenko2020introducing}, we consider a scenario where speakers produce speech utterances that they would like to share with a voice-based service or donate to a public corpus.
As discussed before, a raw speech utterance leaks identifying information embedded in the speaker's voice.
To mitigate this, our goal is to design a speaker anonymization method which takes as input a speech utterance and satisfies the following properties:
\begin{enumerate}
    \item it outputs a speech waveform with the same length as the original speech waveform;
    \item it preserves as well as possible the phonetic and prosodic content of the original utterance (utility);
    \item it conceals as well as possible the identity of the speaker (privacy).
\end{enumerate}
\looseness=-1 Property 1 allows human annotators to listen to the ano\-ny\-mi\-zed speech and annotate it, which is crucial for building useful corpora. The requirement that the speech rate should not be modified comes from the VoicePrivacy challenge \cite{tomashenko2020introducing}. Properties 2 and 3 are conflicting to a certain extent and lead to a classic \emph{privacy-utility trade-off}.

On the one hand, utility can be measured by the performance of an ASR system trained and tested on anonymized utterances. On the other hand, quantifying the level of privacy requires to define a threat model. In this work, we consider adversaries that aim to verify whether a given speaker spoke a target anonymized utterance. We assume that the speaker anonymization method used to anonymize the utterance is public and thus fully known to the adversary. We further assume that adversaries have access to some raw speech utterances from the hypothesized speaker as well as to a large public speech corpus with speaker labels.\footnote{See e.g., the Librispeech data set: \url{https://www.openslr.org/12}}
With the above knowledge, an adversary can anonymize the public corpus with the same method that was used to protect the target utterance, and use the resulting data to train an anonymization-aware ASV system (see Section~\ref{sec:background_speech}). This system can then be deployed to conduct the attack by comparing the target utterance to those from the hypothesized speaker.
This attack scenario corresponds to the strongest attack model introduced in \cite{srivastava2019evaluating} (called ``informed'' adversaries therein).

In this work, we design a novel speaker anonymization method for which we provide both formal and empirical privacy guarantees, in the form of differential privacy and ASV error rates achieved by a concrete adversary, respectively.

\section{Existing Speaker Anonymization Methods}
\label{sec:related_work}

Due to legal and technological awareness in society, privacy-preserving data publishing approaches specific to speech data have recently gained prominence and several methodologies have been proposed for speaker anonymization \cite{tomashenko2020introducing}. In this section, we describe existing techniques and their limitations.

\myparagraph{Speech transformation}, also called voice transformation \cite{stylianou2009voice}, refers to modifications of speech that aim to shift the perceived attributes of an utterance in a certain direction while leaving the linguistic content unchanged. Speech transformation is the simplest kind of speaker anonymization since it does not require large data sets for training machine learning models; instead, it relies on classic signal processing to modify speech parameters. Patino et al.\ \cite{patino2020speaker} presented a speech transformation method for speaker anonymization that alters the spectral envelope, the smooth curve that follows the peaks of the spectrum for any given analysis frame and is governed by the shape of the vocal tract. Specifically, they altered the pole angles of the linear prediction (LP) spectral envelope via the McAdams coefficient \cite{mcadams1984spectral}.  Gupta et al.\ \cite{gupta2020design} improved this work by modifying both the pole angles and the pole radii of the LP spectral envelope. Although these parameter manipulations are perceptually reasonable, speaker information can be easily recovered by machine learning-based attacks, such as training an ASI or ASV system on transformed speech \cite{srivastava2019evaluating}.

\myparagraph{Adversarial training} has been used for speaker anonymization by training neural networks to compute representations of speech that maximize the accuracy of a certain utility task while minimizing the accuracy of speaker identification \cite{srivastava2019privacy}.
For instance, Champion et al.~\cite{champion:hal-02995855} showed that the phonetic features extracted from a neural network trained for automatic speech recognition contain residual speaker information and used speaker adversarial training similar to \cite{srivastava2019privacy} to remove it. A common criticism against adversarial methods is that they do not guarantee protection against other speaker identification attacks than the one implemented by the adversarial branch. In practice, it has been shown experimentally that they also fail against identification attacks following the same architecture as the adversarial branch, due to the fact that they do not generalize well to unseen speakers \cite{srivastava2019privacy}.

\myparagraph{Voice conversion} aims to modify the original speaker's voice (called \emph{source}) such that it sounds like another speaker's voice (called \emph{target}), while leaving the linguistic content unchanged. Bahmaninezhad et al.\ \cite{bahmaninezhad2018convolutional} performed speaker anonymization by converting the original speaker's voice into the average of all voices of the same gender. Pobar and Ipšić \cite{pobar2014online} pre-trained a set of speaker transformations for fixed source-target pairs and identified the source to select one of the corresponding transformations. Yoo et al.\ \cite{yoo2020speaker} presented a many-to-many CycleGAN variational autoencoder-based voice conversion method which encodes the identity of each speaker by a one-hot vector. These methods are hardly applicable in practice: they require many utterances the source speaker to be present in the training set while, in the context of anonymization, the source speaker is usually unknown at training time and limited to a single utterance at test time. Furthermore, except for \cite{bahmaninezhad2018convolutional}, they use real speakers as targets, which can be seen as a form of voice spoofing and raises ethical concerns.

\myparagraph{Speech synthesis} techniques have also been proposed to relax the above requirement. For instance, Justin et al.\ \cite{justin2015speaker} transcribed speech into a diphone sequence and re-synthesized it using a single target speaker. Speech synthesis methods suffer from three limitations. First, they still result in a limited set of target speakers or speaker transformations, which prevents the original speaker from choosing an arbitrary unseen speaker as the target. Second, using a real speaker's voice as the target raises ethical concerns. Third, the conversion of speech into a sequence of discrete tokens as in \cite{justin2015speaker} is error-prone and destroys the prosodic information.
These limitations motivate the goal of converting the original utterance into an arbitrary pseudo-speaker's voice without relying on a transcription step.
This goal can be achieved using x-vector based anonymization, which is considered to be the current state-of-the-art \cite{fangspeaker,tomashenko2020introducing,srivastava2020design,srivastava:hal-03197376,mawalim2020xvector,han2020voice,champion2021study,turner2020speaker,espinoza2020speaker}. The core idea is to extract the sequences of phonetic and prosodic features of the source utterance along with a single x-vector for the whole utterance, and to replace that x-vector by a target x-vector that does not correspond to a real speaker.\footnote{Strategies where the choice of target x-vector does not depend on the input utterance was shown to give best performance in \cite{srivastava2020design}.}
This target x-vector, along with the original phonetic and prosodic features, is provided as input to a neural source-filter (NSF) speech synthesizer \cite{wang2019neural} to produce anonymized speech.

\looseness=-1 In this paper, we choose x-vector based anonymization as our baseline, since it addresses the limitations of speech transformation, adversarial training, voice conversion, and classical speech synthesis techniques outlined above. Yet, it suffers from one remaining limitation: it assumes the identity markers of the source speaker to be concentrated in the x-vector extracted from the utterance, so that replacing it with the target x-vector is sufficient to remove them. In this work, we show that this assumption is incorrect (see Section~\ref{sec:direct_eval}) and provide an effective way to remove residual information about the source speaker and to obtain provable privacy guarantees.

\section{Proposed Approach}
\label{sec:ProposedMethod}

\subsection{Overview}

\begin{figure}[t!]
    \centering
    \includegraphics[width=0.9\columnwidth]{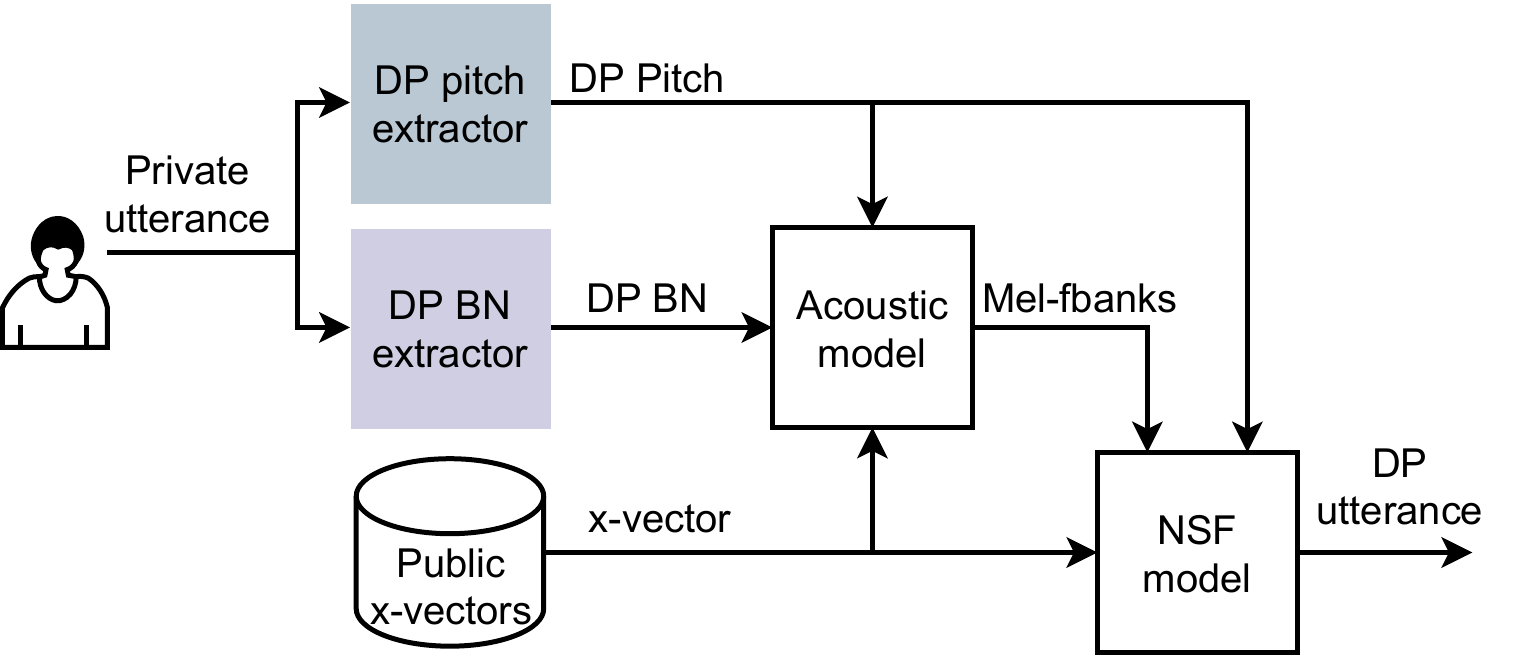}
    \caption{Overview of our proposed speaker anonymization method. Our main contributions are the differentially private pitch and BN feature extractors (shown in color), which make the full pipeline differentially private.}
    \label{fig:BlockDiagram}
\end{figure}

Our speaker anonymization approach is depicted in Figure~\ref{fig:BlockDiagram}.
The overall pipeline is based on x-vector anonymization \cite{fangspeaker,srivastava2020design} (see Section~\ref{sec:related_work}). Pitch and bottleneck (BN) features are first extracted from the input speech.
These features, along with a public speaker embedding (x-vector) that corresponds to a different (pseudo) speaker, are then used to re-synthesize speech using acoustic\footnote{In the context of speech synthesis, the acoustic model generates a sequence of Mel-frequency spectra from the BN and pitch sequences, which is then transformed into a speech waveform by the NSF model. This acoustic model is distinct from that of ASR which, given an input sequence of acoustic features, estimates the corresponding triphone posterior probabilities.} and NSF models.
Note that the x-vector is chosen independently of the input utterance.
Therefore, information about the input speaker can only leak through pitch and BN features. As we will see in our experiments (Figure~\ref{fig:DirectBNUtility}), these features are actually very predictive of speaker identity. Our contribution is to design pitch and BN feature extractors that satisfy differential privacy so as to provably upper bound the amount of residual speaker information embedded in these features while preserving linguistic and prosodic content.\footnote{With a slight abuse of terminology, we will sometimes use \emph{DP pitch} and \emph{DP BN} to refer to features obtained with our DP extractors.} Crucially, the post-processing and composition properties of DP will guarantee that our full pipeline (from the input speech to the anonymized speech) also satisfies DP.

Our DP pitch extractor consists of a conventional pitch estimator followed by an autoencoder network with a Laplace noise layer trained to reconstruct the global pitch dynamics using a custom loss function.
Our DP BN extractor is a deep ASR acoustic model, also with a Laplace noise layer, trained on speech utterances to estimate the corresponding word sequence. We use a public set of annotated speech utterances to train both extractors prior to deployment.
We emphasize that our extractors are quite generic. They may be used in variants of x-vector based speaker anonymization \cite{mawalim2020xvector,champion2021study,turner2020speaker,espinoza2020speaker}, or independently. For instance, we will empirically show that~our DP BN features are sufficient to decode the linguistic content.

In the rest of this section, we
describe our DP pitch and BN feature extractors in detail, and conclude by stating the DP guarantees for our full pipeline.

\myparagraph{Notations.} Let $\x$ be a speech utterance consisting of $K$ time frames. The value of $K$ depends on the duration of the utterance and the chosen frame rate (typically, 10~ms).
The pitch sequence computed from $\x$ is a non-negative 1-dimensional sequence of length $K$, which we denote by a vector ${\p \in \mathbb{R}_+^{K}}$. The BN features extracted from $\x$ are an $M$-dimensional sequence of length $K$ that we denote by a matrix $\bnm = [\bn_1,\dots,\bn_K]^\top \in \mathbb{R}^{K\times M}$ where each $\bn_k\in\mathbb{R}^M$. We denote by $W$ the ground truth text transcription of $\x$.

\myparagraph{Public data.} We assume that we have access to a \emph{public} data set $\mathcal{X}=\{(\x_i, W_i)\}_{i=1}^N$ of $N$ annotated speech utterances to train our DP feature extractors. This training data set must be disjoint from the (private) data used at deployment, i.e., speakers have to be different in both datasets.

\subsection{Differentially Private Pitch Extractor}
\label{subsec:dp_pitch}

As mentioned earlier, the global dynamics of the pitch sequence $\p$ for an utterance $\x$ conveys prosodic information, while its local variations are more specific to each speaker \cite{adami2003moeling,peskin2003using,dehak2007modeling,mary2008extraction}.
We aim to learn a DP autoencoder $\map$ which takes as input a raw pitch sequence $\p$ computed by a conventional pitch estimator, and outputs a perturbed pitch sequence $\p^{\texttt{DP}}$ of the same length in which the identity information has been removed while most of the prosodic information is preserved.
An obvious approach to obtain a DP autoencoder is to rely on input perturbation, i.e., to add Laplace noise directly to the raw pitch $\p$. However,
this baseline largely destroys the time correlations that are indicative of prosody elements that we wish to preserve, as we show in our experiments.

\begin{figure}[t!]
    \centering
    \includegraphics[width=0.90\columnwidth]{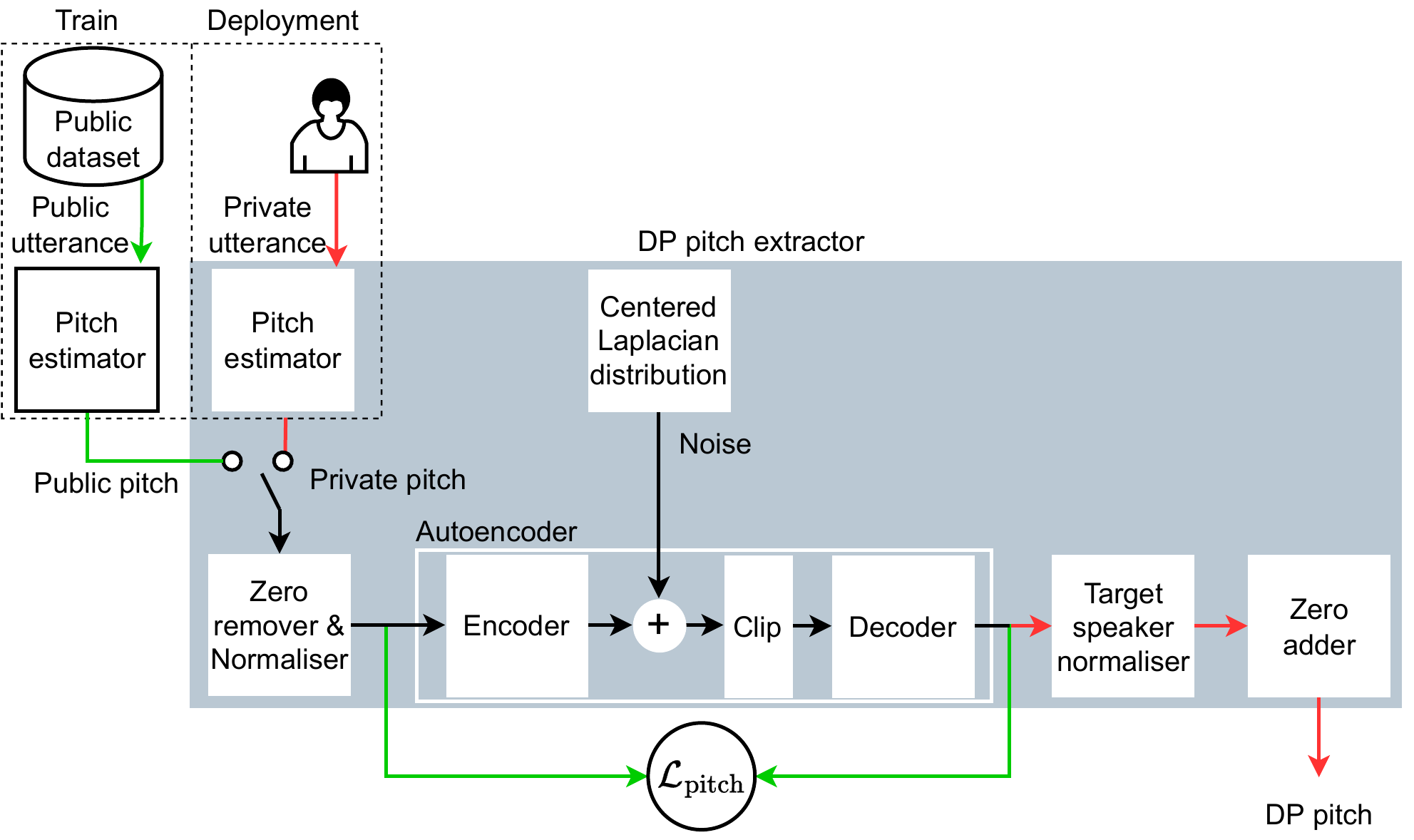}
    \caption{Proposed DP pitch extractor. The convolutional autoencoder with a noise layer is trained using public pitch sequences and subsequently used to generate perturbed pitch sequences from private pitch sequences in a differentially private fashion. Black arrows show paths that are common to both training and deployment, while green and red arrows apply only to training or deployment, respectively.}
\label{fig:BlockDiagramDPF0AE}
\end{figure}

Instead, we propose to learn a deep convolutional autoencoder with a noise layer. Below, we describe the architecture of our autoencoder, how it is trained, and finally how it can be deployed to anonymize pitch sequences.
The block diagram of our complete DP pitch extractor is shown in Figure~\ref{fig:BlockDiagramDPF0AE}.

\myparagraph{Autoencoder architecture.} We propose to define $\map=\dec\circ\noise_p\circ\enc$ as a fully convolutional autoencoder
composed of an encoder $\enc$, a noise layer $\noise_p$ and a decoder $\dec$. 
This encoder-decoder architecture, inspired by \cite{shamsabadi2021foolhd}, has two important benefits in our context.
First, a fully convolutional architecture enables us to deal with variable-length input and output sequences as the shape and size of the weights of each convolutional layer (kernel) are not affected by the size of the input and output of that layer. Second, convolutional layers are suitable to capture time dependencies in pitch sequences.

The encoder $\enc$ maps an input pitch $\p\in\mathbb{R}^K$ to a latent representation $\h=\enc(\p) \in [0,1]^{C\times K}$
through 3 convolutional layers (each with $C$ channels) with sigmoid activation functions.

\looseness=-1 In order for the autoencoder $\map$ to satisfy $\varepsilon$-DP for a given $\varepsilon>0$, the encoder is followed by a noise layer $\noise_p$ which adds centered Laplace noise to each entry of the latent representation $\h$ to generate a perturbed version $\h^{\texttt{DP}}\in \mathbb{R}^{C\times K}$:
\begin{equation}
\h^{\texttt{DP}} = \noise_p(\h)= \h + \text{Lap}(\Delta_1(\enc)/\varepsilon),
\end{equation}
where $\Delta_1(\enc)=\max_{\p,\p'}\|\enc(\p)-\enc(\p')\|$ is the $\ell_1$-sensitivity of $\enc$.
While tightly bounding the sensitivity of neural networks can be challenging in general \cite{papernot2020tempered}, here the use of the sigmoid activation allows us to easily bound $\Delta_1(\enc)$ since each entry of $\h$ belongs to $[0,1]$:
\begin{equation}
    \Delta_1(\enc)=C \times K \times 1 = C K.
\end{equation}
This bound is tight enough in practice for the Laplace noise injected to the features not to be detrimental to utility, as long as the value of $\varepsilon$ remains reasonable (away from zero).

Finally, the decoder $\dec$ takes as input the perturbed latent representation $\h^{\texttt{DP}}$, deterministically clips each of its entries back to $[0,1]$ (which we found to help training to converge), and decodes it into a perturbed pitch sequence  $\p^{\texttt{DP}}=D(\h^{\texttt{DP}})\in\mathbb{R}^K$ through 3 convolutional layers (2 $C$-channel layers with sigmoid activation followed by 1 layer with linear activation).

\begin{figure}[t]
\centering
\begin{tikzpicture}
\scriptsize
\begin{axis}[cycle list name=color list,
      axis lines=left, 
      width=8.5cm,
      height=3.5cm,
      ymin=80,
      ymax=140,
enlarge y limits=0.01,
      enlarge x limits=0.02,
      smooth,
      xlabel={\scriptsize Frame},
      ylabel={\scriptsize Pitch (Hz)},
    ytick={0,20,40,60,80, 100, 120, 140,160},
    xtick={40,60,80, 100, 120, 140,160,180,200,220},
    y label style={at={(axis description cs:-.07,.5)},anchor=south},
            xtick style={draw=none},
        ytick style={draw=none},
      ]
\addplot+[black] table [x=x, y=y]{./tikz/org.txt};\label{fig:orgP}
\addplot+[blue] table [x=x, y=y]{./tikz/dp10.txt};\label{fig:dpP10}
\addplot+[red] table [x=x, y=y]{./tikz/dp1.txt};\label{fig:dpP1}
\end{axis}

\end{tikzpicture}
\caption{\looseness=-1 Visualization of the original (non-private) pitch sequence ($\ref{fig:orgP}$) and noisy reconstructed pitch sequences obtained with our approach for $\varepsilon=10$ ($\ref{fig:dpP10}$) and $\varepsilon=1$ ($\ref{fig:dpP1}$). In general, our approach preserves the global dynamics of the original pitch sequence thanks to our correlation-based loss. }
\label{fig:VisPitch}
\end{figure}
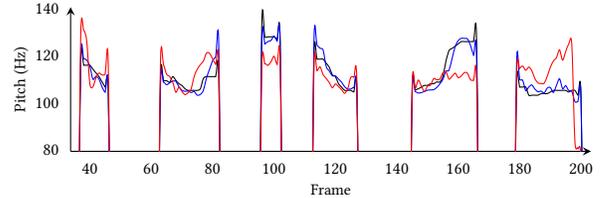

\myparagraph{Training phase.} We train our autoencoder on a set of raw pitch sequences $\{\p_i \in \mathbb{R}^{K_i}\}_{i=1}^{N}$ computed from the speech waveforms $\x_i$ in the public data set $\mathcal{X}$ (using for instance the YAAPT estimator). The pitch sequences are pre-processed as follows. First, zero values are removed. Indeed, these values indicate silence or unvoiced phonemes and must be kept to zero so that silence remains silence, and every unvoiced phoneme remains the same unvoiced phoneme. For instance, replacing the zero pitch on phoneme \textit{/p/} by a nonzero pitch would transform it into a \textit{/b/}, which would harm utility. This operation does not affect privacy,
since these values are equal to zero for all speakers and therefore do not convey identity information. It also makes it possible to account for variations of pitch across successive voiced phonemes. Table~\ref{tab:PitchStats} shows statistics on the length of pitch sequences and the proportion of zero values in a data set used in our evaluation, and Figure~\ref{fig:VisPitch} shows an example of raw pitch sequence. Finally, since pitch differs in range across speakers, the last step of pre-processing is to normalize each sequence to have zero mean and unit variance.

\looseness=-1 To preserve the prosodic content in the reconstructed pitch, we propose to train the autoencoder by minimizing the following loss:
\begin{equation}
\label{eq:loss}
    \mathcal{L}_{\text{pitch}}=1-\sum_{i=1}^N\texttt{Corr}(\p_i,\p_i^{\texttt{DP}}),
\end{equation}
\looseness=-1 where $\texttt{Corr}(\cdot,\cdot)$ is the Pearson correlation coefficient. In contrast to the standard mean squared error loss, maximizing correlations between the original and reconstructed pitch makes reconstruction errors in the local variations of the pitch (which tend to be more speaker-specific) more costly than errors in the global dynamics of the sequence (which convey the global prosodic content of the utterance). As the autoencoder \emph{must} suffer reconstruction errors due to the addition of noise, it will learn to preserve global dynamics as much as possible while sacrificing local variations, which is the desired behavior for speaker anonymization.
This is illustrated in Figure~\ref{fig:VisPitch}.

\myparagraph{Deployment phase.}
Once the autoencoder $\mathcal{A}$ has been trained, we can use it to generate perturbed versions of the pitch sequence of any private utterance $\x$. Similarly to the training phase, we compute the pitch sequence $\p$ from $\x$, remove the zeros, normalize it, and push it to the autoencoder to obtain a perturbed pitch sequence $\p^{\texttt{DP}}$. We then normalize $\p^{\texttt{DP}}$ to match the mean $\mu_\text{target}$ and variance $\sigma_\text{target}$ of the pitch of a target (pseudo) speaker, where $\mu_\text{target}$ and $\sigma_\text{target}$ are computed over a public set of utterances from the target speaker.
In the context of the full speaker anonymization pipeline of Figure~\ref{fig:BlockDiagram}, this normalization (which we call \emph{pitch conversion})
makes the mean and variance of the perturbed pitch consistent with the choice of the target x-vector.
Finally, we add the zero values back in their original positions in the sequence.

\myparagraph{Privacy guarantees.}
By the Laplace mechanism (Definition~\ref{defn:laplace-mech}), $\noise_p\circ\enc$ satisfies $\varepsilon$-DP, and so does the autoencoder $\map=\dec\circ\noise_p\circ\enc$ by the post-processing property of DP.

\begin{table}[t] 
    \centering 
    \footnotesize
    \caption{\looseness=-1 YAAPT pitch statistics on the dev\_clean subset of LibriSpeech.} 
       \begin{tabular}{lcccc} 
    \toprule 
               &  \textbf{Min} & \textbf{Max}   & \textbf{Avg}  & \textbf{Std} \\ 
     \midrule                
        Length $K$ &  147 & 3261  & 743  & 493 \\
Non-Zeros  & 24\% & 76\%  & 53\% & 8\%\\
    \bottomrule
    \end{tabular}
    \label{tab:PitchStats}
\end{table}

\subsection{Differentially Private BN Extractor}

\begin{figure}[t!]
    \centering
    \setlength{\tabcolsep}{2pt} 
    \includegraphics[width=0.95\columnwidth]{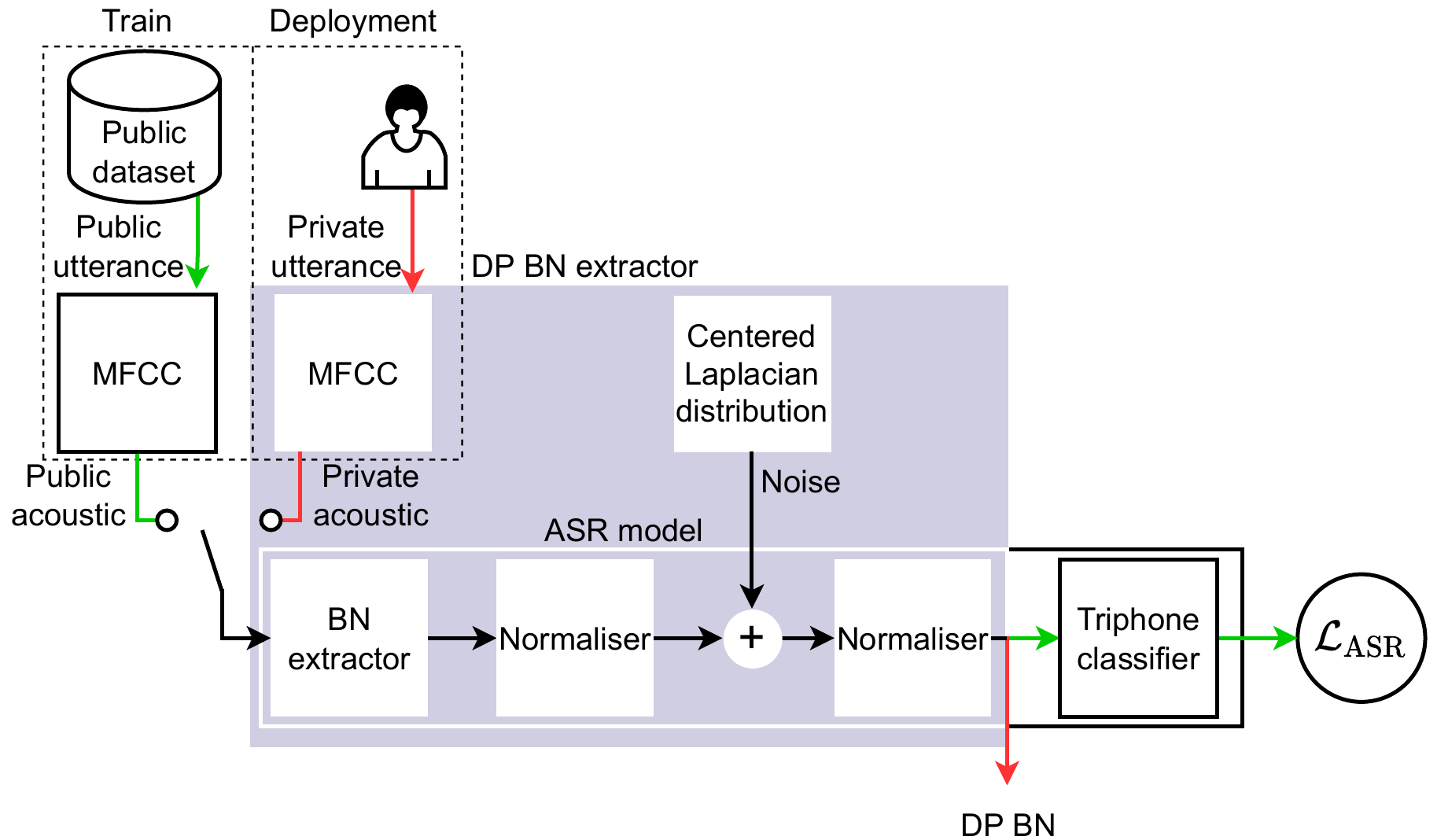}
    \caption{\looseness=-1 Proposed DP BN extractor. The ASR acoustic model with a noise layer is trained on public utterances and subsequently used to generate perturbed BN features from private utterances in a DP fashion. Black arrows show paths that are common to both training and deployment, while green and red arrows apply only to training or deployment, respectively.}
    \label{fig:DPBN}
\end{figure}

\looseness=-1 We now turn to the BN features, which are phonetic features that should be sufficient to decode the linguistic information. BN features are typically obtained as an intermediate layer of an ASR acoustic model. However, traditional BN features also contain residual speaker information as shown in our experiments (see Section~\ref{sec:direct_eval}). We propose to address this issue by adding a noise layer to the acoustic model, similarly to the approach used for pitch.  The block diagram of our complete DP BN extractor is shown in Figure~\ref{fig:DPBN}.

\myparagraph{ASR model architecture.} We adapt here the widely used ASR acoustic model architecture and sequence-discriminative training criterion proposed in \cite{povey2016purely}.\footnote{We note that our approach can be easily applied to other acoustic models.} We view the ASR acoustic model $\M=\T\circ\noise_B\circ\B$ as three sequential parts: a BN extractor $\B$, followed by a noise layer $\noise_B$, and finally a triphone classifier $\T$.
The BN extractor $\B$ takes as input a sequence of acoustic features $\mathbf{O}\in\mathbb{R}^{K\times A}$ extracted from
a speech utterance $\x$ with $K$ frames and outputs a sequence of BN features $\bnm = [\bn_1,\dots,\bn_K]^\top \in \mathbb{R}^{K\times M}$:
\begin{equation}
    \bnm = \B(\mathbf{O}).
\end{equation}
The acoustic features $\mathbf{O}$ are the concatenation of 40-dimensional MFCCs, which are the most popular spectral features for speech processing, and 100-dimensional i-vectors \cite{saon2013speaker} which help the acoustic model adapt to different speakers. Therefore, the per-frame dimensionality of these features is $A=140$. The BN extractor $\B$ is composed of 17 factorized time delay neural network layers \cite{Povey2018}, which perform one-dimensional convolution operations to learn the temporal context present in the acoustic feature sequence $\mathbf{O}$. 

We now describe our noise layer $\noise_B$, which we use to hide speaker information and achieve differential privacy. Each frame-level BN feature vector $\bn_k$ is $M$-dimensional with $M=256$. Due to this high dimensionality, we enforce $\varepsilon$-DP at the frame level (from which we can deduce DP guarantees at the utterance level, as explained at the end of this section).
Note that each frame-level BN feature vector $\bn_k$ is not normalized. Therefore, our noise layer $\noise_B$ first normalizes each $\bn_k$ to have unit $\ell_1$-norm and then adds Laplace noise to their entries to generate a sequence of perturbed BN features $\bnm^{\texttt{DP}} = [\bn^{\texttt{DP}}_1,\dots,\bn^{\texttt{DP}}_K]^\top \in \mathbb{R}^{K\times M}$:
\begin{equation}
\label{eq:perturbed-BN}
\bnm^{\texttt{DP}} = \noise_B(\bnm)=
[\noise_b(\bn_1),\dots,\noise_b(\bn_K)]^\top,
\end{equation}
\looseness=-1 where $\noise_b(\bn)=\text{norm}_1(\text{norm}_1(\bn) + \text{Lap}(2/\varepsilon))$ with $\text{norm}_1(\mathbf{c})=\mathbf{c}/\|\mathbf{c}\|_1$. The scale of the centered Laplace noise comes from the application of the Laplace mechanism (Definition~\ref{defn:laplace-mech}), where the $\ell_1$-sensitivity of the normalized frame-level BN features is bounded by $\max_{\mathbf{b},\mathbf{b’}}\|\mathbf{b}-\mathbf{b’}\|_1 \leq 2$ based on the fact that we normalize each BN feature vector $\bn$ to have $\|\mathbf{b}\|_1=1$ and by triangle inequality $\|\mathbf{b}-\mathbf{b’}\|_1 \leq \|\mathbf{b}\|_1+\|\mathbf{b’}\|_1=2$.

Note that we post-normalize the perturbed BN features to have unit $\ell_1$-norm, as we found this to improve training convergence.

Finally, the triphone classifier $\T$ takes the sequence of perturbed BN features $\bnm^{\texttt{DP}}$ as input and outputs the corresponding triphone log-posterior probabilities $\{P(S_k|\bnm^{\texttt{DP}})\}_{k=1}^N$
which represent the phonetic content of $\x$ (see Section~\ref{sec:background_speech}). We refer to \cite{povey2016purely} for details on the architecture of $\T$.

\myparagraph{Training phase.}
\label{para:asr-training}
Our ASR acoustic model $\M$ is trained on acoustic features $\{\mathbf{O}_i\}_{i=1}^N$ extracted from the utterances $\{\x_i\}_{i=1}^N$ in the public annotated data set $\mathcal{X}$ and their corresponding transcriptions $\{W_i\}_{i=1}^N$.
Our goal is that the model learns to keep sufficient information in the BN features to predict the linguistic content, while other information (in particular, speaker identity) is naturally lost due to noise addition.
To this end, we minimize a cost function ${\mathcal{L}_\text{ASR}=\mathcal{L}_\text{MMI} + 0.1\cdot\mathcal{L}_\text{CE}}$ composed of two terms. The dominant term, $\mathcal{L}_\text{MMI}$, is the lattice-free maximum mutual information (LF-MMI) \cite{povey2016purely} cost which aims to maximize the posterior probability of the ground truth word sequence $W_i$:
\begin{equation}
    \mathcal{L}_\text{MMI} = - \sum_{i=1}^N \log \frac{P(\mathbf{O}_i|W_i) P(W_i)}{\sum_{W'} P(\mathbf{O}_i|W') P(W')}.
\end{equation}
The numerator is the joint likelihood of the acoustic features $\mathbf{O}_i$ and the ground truth word sequence $W_i$, while the denominator is the likelihood of the acoustic features marginalized over all possible word sequences. The numerator is computed by summing over all triphone sequences corresponding to 
$
W_i:P(\mathbf{O}_i|W_i)=\sum_{S_i,N_i} P(\mathbf{O}_i|S_i) P(S_i|N_i) P(N_i|W_i)
$
where $P(\mathbf{O}_i|S_i)\propto\prod_{k=1}^{K_i}P(S_{i,k}|\bnm^{\texttt{DP}}_i)/P(S_{i,k})$, and $P(S_{i,k})$, $P(S_i|N_i)$, $P(N_i|W_i)$ and $P(W_i)$ are fixed as explained in Section \ref{sec:background_speech}. The numerator is computed in a similar way, except that the (intractable) sum over all possible word sequences with a word-level language model is approximated by a (tractable) sum over all possible phoneme sequences with a phoneme-level language model.
The second term $\mathcal{L}_\text{CE}$ of the cost function is the frame-level cross-entropy loss between true and estimated triphone states, which acts as a regularizer \cite{povey2016purely}:
\begin{equation}
    \mathcal{L}_\text{CE} = -\sum_{i=1}^N \sum_{k=1}^K \log P(S_{i,k}|\bnm^{\texttt{DP}}_i).
\end{equation}

\myparagraph{Deployment phase.}
Once the ASR acoustic model $\M=\T\circ\noise_B\circ\B$ has been trained, we can use it to generate a sequence of perturbed BN features $\bnm^{\texttt{DP}}=\noise_B\circ\B(\mathbf{O})$ for any private utterance $\x$ with acoustic features $\mathbf{O}$.

\myparagraph{Privacy guarantees.} By the Laplace mechanism, the frame-level mechanism $\noise_b$ satisfies $\varepsilon$-DP. Note that this frame-level guarantee can be converted into a rigorous utterance-level guarantee using the composition property of DP. In particular, the BN extractor $\noise_B\circ\B$ satisfies $\varepsilon'$-DP with $\varepsilon'=K\varepsilon$ for an utterance of length $K$.
The utterance-level bound is thus looser by a factor of the utterance length $K$. Yet, we will see in our experiments that even when the \emph{analytical} utterance-level privacy guarantee is rather weak (i.e., large $\varepsilon'$ due to large values of $K$), noise addition still has a significant effect on the \emph{practical} protection of utterances against strong speaker identification and speaker verification attacks.

\subsection{Privacy Guarantees for the Full Pipeline}

Our full speaker anonymization pipeline (from an utterance to its anonymized version) can be seen as a composition of two DP mechanisms (the DP pitch and DP BN extractors) followed by post-processing steps that do not depend on the input utterance, see Figure~\ref{fig:BlockDiagram}. Therefore, we can directly obtain DP guarantees for the full pipeline by composing the DP guarantees of the DP pitch and BN extractors.

Formally, let the DP pitch extractor satisfy $\varepsilon_1$-DP and the DP BN extractor satisfy $\varepsilon_2$-DP (at the frame level). Then, by the composition property of DP, the combined output of these extractors satisfy $(\varepsilon_1+K\varepsilon_2)$-DP for an utterance of length $K$. From the robustness to post-processing property of DP, the full pipeline also satisfies $(\varepsilon_1+K\varepsilon_2)$-DP.

\section{Empirical Validation Setup}
\label{sec:val}
In complement to the analytical privacy guarantees obtained in Section~\ref{sec:ProposedMethod}, we design an empirical evaluation of the privacy and utility of our approach, and compare against the state-of-the-art speaker anonymization scheme.

\subsection{General Objectives}
Our DP pitch and BN extractors aim to remove the residual speaker identity information from the prosodic and linguistic attributes of an utterance, which are then used in our DP speaker anonymization pipeline of Figure~\ref{fig:BlockDiagram} to output utterances with rich prosodic and linguistic attributes. Therefore, our experiments consider the following  major dimensions:
\begin{enumerate}
    \item How much identity information is retained within the original pitch, BN features and anonymized utterances?
    \item How well can our approach remove this residual speaker information from pitch, BN features and utterances, and protect against concrete attacks?
    \item How does our approach affect the utility of utterances, for ASR training and inference?

\end{enumerate}

\subsection{Data Set}
We work with the LibriSpeech data set \cite{panayotov2015librispeech} used in the VoicePrivacy challenge \cite{vpc2022csl}. It contains about 1,000 hours of English speech derived from audiobooks.  LibriSpeech is partitioned into five subsets: train\_clean\_100, train\_clean\_360, train\_other\_500, dev\_clean and test\_clean. We also use a subset of \mbox{LibriTTS} \cite{libritts} (a text-to-speech dataset derived from Librispeech) to train the speech synthesis component of the systems. We refer to Appendix~\ref{app:data_set} for details on these subsets.
Following the VoicePrivacy challenge \cite{vpc2022csl}, we consider the train\_clean\_100, train\_clean\_360, train\_other\_500 and \mbox{LibriTTS} subsets as public data set for training our DP extractors, the speech synthesis models, the attacks and the ASR models used for utility evaluation, and the test\_clean subset as the private utterances seen in the deployment phase.

\subsection{Speaker Anonymization Methods under Comparison}

We compare our differentially private speaker anonymization approach (\DP) with the state-of-the-art x-vector based method (\Anon) \cite{fangspeaker,srivastava2020design}.
To analyze the impact of each of our DP extractors separately, we also consider two partial instantiations of our method: i) \AnonDPBN, a modification of \Anon where the BN extractor is replaced by our DP BN extractor; ii) \AnonDPPitch, a modification of \Anon where the pitch extractor is replaced by our DP pitch extractor followed by pitch conversion. Note that the original \Anon method does not perform pitch conversion. Therefore, when evaluated against \AnonDPPitch and \DP, we enhance \Anon with pitch conversion (\AnonPC) for a fair comparison.
The different variants are summarized in Table~\ref{tab:Dpversions}.

\begin{table}[t] 
    \centering 
    \footnotesize
\caption{Competing anonymization methods, including different variants of our approach.
KEYS -- Anon: Anonymization; PC: Pitch Conversion; DP: Differential Privacy.} 
       \begin{tabular}{lcccc} 
    \toprule 
         \textbf{Name}      &  \textbf{DP pitch} & \textbf{PC} & \textbf{DP BN} & \textbf{x-vector}\\ 
              &  \textbf{extractor} &  & \textbf{extractor} & \textbf{anon} \\ 
     \midrule                
         \Anon &  - & - & - & \checkmark\\
         \AnonPC  &  -  & \checkmark & - & \checkmark\\
         \AnonDPBN (ours) &  - & - & \checkmark & \checkmark\\
         \AnonDPPitch (ours) &  \checkmark & \checkmark & -& \checkmark\\
        \DP (ours) & \checkmark & \checkmark & \checkmark & \checkmark\\
    \bottomrule
    \end{tabular}
\label{tab:Dpversions}
\end{table}

All systems are trained on feature sequences with a frame rate of 10~ms. For pitch estimation, we use YAAPT \cite{kasi2002yet}.
Our proposed DP pitch extractor is implemented in PyTorch and trained on train\_clean\_100.
For each system, the BN extractor is trained using the Kaldi toolkit \cite{povey2011kaldi} on the combined train\_clean\_100 and train\_clean\_500 data subsets as in \cite{tomashenko2020introducing}, and the speech synthesis component (acoustic and NSF models) is trained on the \mbox{LibriTTS} subset, as in \cite{tomashenko2020introducing}.
In all systems and consistently with Figure~\ref{fig:BlockDiagram}, the target x-vector assigned to a private utterance is selected independently of that utterance, preventing any leakage of speaker identity information from this step.\footnote{Previous work alternatively considered speaker-level assignment (i.e., same target x-vector used for all utterances of a given speaker) \cite{vpc2022csl,srivastava:hal-03197376}, but this introduces an undesirable dependence on the speaker identity. In Appendix~\ref{app:utterance_vs_speaker}, we show that speaker-level assignment does not provide better empirical protection than utterance-level assignment, and discuss the choice of assignment strategy from the point of view of the attacker.} We use a variant of the approach proposed by \cite{srivastava2020design} where the set of public x-vectors is first clustered, then a dense cluster is selected, and finally the target x-vector is chosen as the average of randomly selected x-vectors from that cluster.\footnote{The averaging step ensures that the target x-vector corresponds to a pseudo speaker, not a real speaker.} Additional details on this selection strategy and on training the speaker anonymization systems are in Appendix~\ref{app:anon_systems}.

\subsection{Attacks}
In addition to the rigorous analytical guarantees of our approach based on the DP analysis, we provide an empirical understanding of the privacy benefits of our approach. We quantify the empirical protection provided by the different speaker anonymization methods against concrete attacks that aim to re-identify a known speaker or to link together two utterances from the same speaker. The adversaries have full knowledge of the anonymization scheme that was applied (including its parameters), and have access to a large public speech corpus (with speaker identities) that they can anonymize using the targeted scheme to train their attack. They also have access to ``enrollment'' utterances from the target speaker.
This corresponds to the strongest attack model that has been considered in speaker anonymization~\cite{srivastava2019evaluating}, allowing us to perform an empirical privacy evaluation closer to the worst-case (maximum privacy leakage).

\looseness=-1 First, we perform speaker re-identification attacks directly on (utterance-level) pitch and BN features. Similar to \cite{srivastava2019privacy}, the attack is based on an ASI system which follows the classical speaker classification architecture \cite{snyder2018x} except that, instead of MFCCs as inputs, it is trained on pitch or BN features (with or without DP depending on the anonymization method under attack) extracted from the train\_clean\_360 data subset. This subset contains 921 speakers and is divided into train/validation/test splits such that 80\% utterances of each speaker are used for training, 10\% for validation and 10\% for testing. The ASI system is trained over the train split, while the validation set is used for monitoring the generalization performance and early stopping in case of convergence. The adversary then uses the trained ASI system to predict the speaker of unseen utterances in the test split, among the known identities from the training set.

Second, we perform speaker linkage attacks on anonymized utterances. In line with the evaluation of the VoicePrivacy challenge \cite{tomashenko2020introducing}, the attack is based on an ASV system (see Section~\ref{sec:background_speech}) that follows the standard setup in Kaldi \cite{povey2011kaldi}.
The ASV system (both x-vector extractor and PLDA) is trained on the train\_clean\_360 data subset, anonymized with the anonymization method under attack.
Finally, the adversary uses the trained ASV system to compute a log-likelihood ratio score between a trial and an enrollment utterance, and decides whether they are from the same speaker by comparing the score with a threshold. The choice of this threshold affects the ratio between the false acceptance rate and the false rejection rate of the attack (see Section~\ref{sec:metrics}).
More details on the implementation of these attacks can be found in Appendix~\ref{app:attacks}.

In line with previous DP studies \cite{DP_eval,jagielski2020auditing,nasr2021adversary}, we expect that our approach may give better protection against such concrete attacks than what the analytical $\varepsilon$-DP guarantee suggests, because DP is a worst-case guarantee and enforces a stronger notion of privacy than concealing the speaker identity.

\subsection{Performance Measures}
\label{sec:metrics}
\myparagraph{Empirical privacy measures.} We define empirical privacy measures based on the performance of the above attacks.

The empirical privacy $\text{P}_\text{ASI}$ achieved by pitch and BN extractors is measured by the error of the speaker re-identification attack, i.e., the proportion of utterances which are \emph{not} assigned to the correct speaker by the ASI system, which varies between 0\% (worst privacy) and 100\% (best privacy). We report $\text{P}_\text{ASI}$ over the (unseen) test split.

The empirical privacy achieved by a full speaker anonymization pipeline is measured by the equal error rate (EER) $\text{P}_\text{ASV,e}$ and unlinkability $\text{P}_\text{ASV,l}$ of the speaker linkage attack. These two metrics are standard in speaker anonymization \cite{Maouche2020a}. The EER is equal to the false acceptance rate and the false rejection rate at the threshold for which these two rates are equal \cite{bimbot2004tutorial}, and it varies between 0\% (worst privacy) and 50\% (best privacy). In constrast, the unlinkability does not depend on the choice of a threshold: it measures the amount of overlap between the \emph{distributions} of same-speaker and different-speaker scores. It is equal to 1 minus the linkability metric in \cite{gomez2017general,Maouche2020a} and varies between 0 (worst privacy) and 1 (best privacy). The trial and enrollment sets used to compute $\text{P}_\text{ASV,e}$ and $\text{P}_\text{ASV,l}$ are constructed from the test\_clean subset in the same way as done in the VoicePrivacy challenge \cite{tomashenko2020introducing}.

\myparagraph{Utility measure.} 
We quantify the preservation of linguistic content of a speaker anonymization scheme by the accuracy of an ASR system trained and evaluated on anonymized utterances. The empirical utility $\text{U}_\text{ASR}$ is equal to 100 minus the word error rate (WER) of the ASR system, i.e., the percentage of word substitutions, deletions, and insertions compared to the number of words in the ground truth transcriptions. 
We train the evaluation ASR system on the train\_clean\_360 data subset following the Kaldi recipe for LibriSpeech \cite{panayotov2015librispeech}, see Appendix~\ref{app:asr} for details.

In summary, the larger $\text{P}_\text{ASI}$, $\text{P}_\text{ASV,e}$ and $\text{P}_\text{ASV,l}$, the greater the privacy, and the larger $\text{U}_\text{ASR}$, the greater the utility.

\section{Empirical Results}

\looseness=-1 Our direct evaluation of pitch and BN features in Section~\ref{sec:direct_eval} shows that DP improves the empirical privacy of pitch and BN features with negligible effects on their utility.  Our systematic study in Section~\ref{sec:SeperatePitchBN}  demonstrates the advantages of plugging our DP pitch and DP BN extractor \emph{separately}, and above all \emph{simultaneously}, on the privacy of the utterances over the previous state-of-the-art approach.

\subsection{Privacy and Utility of Pitch and BN}
\label{sec:direct_eval}

\begin{figure}[t!]
\pgfplotsset{
    every non boxed x axis/.style={},
    boxplot/every box/.style={solid,ultra thin,black},
    boxplot/every whisker/.style={solid,ultra thin,black},
    boxplot/every median/.style={solid,very thick, red},
}
\pgfplotstableread{tikz/DirectBN/EER.txt}\eer
\pgfplotstableread{tikz/DirectPitch/EER.txt}\eerp
\begin{tikzpicture}
\node[inner sep=0pt] (whitehead) at (0.5,-0.2)
{\footnotesize Baseline};
\node[inner sep=0pt] (whitehead) at (2,-0.2)
{\footnotesize DP BN};
    \footnotesize
        \begin{axis}[
        axis x line=bottom,
        axis y line=left,
        width=4.95cm,
        height=3.5cm,
        boxplot/draw direction=y,
        xtick={1,2.5,3.5,4.5,5.5},
        ymin=0,ymax=100,
        ytick={0,20,40,60,80,100},
        title=BN features,
        yticklabels={0,20,40,60,80,100},
        xticklabels={\vphantom{.1cm}$\varepsilon$=$\infty$,$\varepsilon$=100,~$\varepsilon$=10,$\varepsilon$=1,$\varepsilon$=0.5},
        xticklabel style={yshift=-10pt},
        ylabel={$\text{P}_\text{ASI}$ (\%)},
        y label style={at={(axis description cs:-.15,.5)},anchor=south},
        xtick style={draw=none},
        ytick style={draw=none},
]
        \addplot+[boxplot, boxplot/draw position=1,mark=*, mark options={black,scale=0},boxplot/box extend=0.5,boxplot/every median/.style={solid,very thick, black}] table[y=einf]{\eer};
\addplot+[boxplot, boxplot/draw position=2.5,mark=*, mark options={red,scale=0},boxplot/box extend=0.5] table[y=e100]{\eer};
\addplot+[boxplot, boxplot/draw position=3.5,mark=*, mark options={red,scale=0},boxplot/box extend=0.5] table[y=e10]{\eer};
\addplot+[boxplot, boxplot/draw position=4.5,mark=*, mark options={red,scale=0},boxplot/box extend=0.5] table[y=e1]{\eer};
\addplot+[boxplot, boxplot/draw position=5.5,mark=*, mark options={red,scale=0},boxplot/box extend=0.5] table[y=e05]{\eer};
        \end{axis}
\end{tikzpicture} 
        \begin{tikzpicture}
\node[inner sep=0pt] (whitehead) at (0.5,-0.2)
{\footnotesize Baseline};
\node[inner sep=0pt] (whitehead) at (1.95,-0.2)
{\footnotesize DP Pitch};
    \footnotesize
        \begin{axis}[
        axis x line=bottom,
        axis y line=left,
        width=4.6cm,
        height=3.5cm,
        boxplot/draw direction=y,
        xtick={1,2.5,3.5,4.5,5.5},
        ymin=0,ymax=100,
title=Pitch,
xticklabels={\vphantom{.1cm}$\varepsilon$=$\infty$,$\varepsilon$=100,~$\varepsilon$=10,$\varepsilon$=5,$\varepsilon$=1},
        xticklabel style={yshift=-10pt},
        ylabel={$\text{P}_\text{ASI}$ (\%)},
        y label style={at={(axis description cs:-.15,.5)},anchor=south},
        xtick style={draw=none},
        ytick style={draw=none},
]
        \addplot+[boxplot, boxplot/draw position=1,mark=*, mark options={black,scale=0},boxplot/box extend=0.5,boxplot/every median/.style={solid,very thick, black}] table[y=einf]{\eerp};
\addplot+[boxplot, boxplot/draw position=2.5,mark=*, mark options={red,scale=0},boxplot/box extend=0.5] table[y=e100]{\eerp};
\addplot+[boxplot, boxplot/draw position=3.5,mark=*, mark options={red,scale=0},boxplot/box extend=0.5] table[y=e10]{\eerp};
\addplot+[boxplot, boxplot/draw position=4.5,mark=*, mark options={red,scale=0},boxplot/box extend=0.5] table[y=e5]{\eerp};
\addplot+[boxplot, boxplot/draw position=5.5,mark=*, mark options={red,scale=0},boxplot/box extend=0.5] table[y=e1]{\eerp};
\end{axis}
\end{tikzpicture}
    \caption{Empirical privacy of \textcolor{black}{original BN features} versus \textcolor{red}{our proposed DP BN features} (left) and \textcolor{black}{original pitch} versus \textcolor{red}{our proposed DP pitch} (right) for different privacy budgets $\varepsilon$. Empirical privacy is assessed by the test error ($\text{P}_\text{ASI}$) of a speaker re-identification attack trained directly on BN features (left) and pitch (right) from the train\_clean\_360 split (921 speakers). Our DP extractors significantly reduce speaker information present in BN feature and pitch.}
\label{fig:DirectBN}
\end{figure}
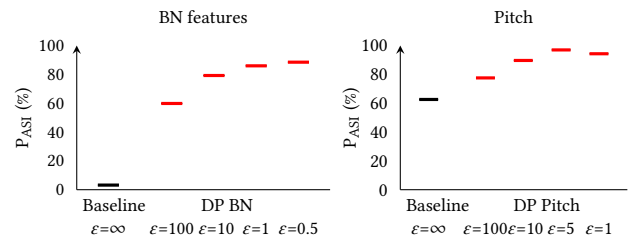

\myparagraph{How much speaker identity information is retained withing pitch and BN features?}
\label{sec:leakage}
Figure~\ref{fig:DirectBN} shows the empirical privacy of BN features (left) and pitch (right) in terms of the test error $\text{P}_\text{ASI}$ of speaker re-identification attacks (ASI systems trained directly on these features). 
First of all, the results clearly demonstrate that original pitch and BN features do retain a lot of speaker information, violating a key assumption supporting previous x-vector based speaker anonymization approaches. Note that BN features contain more identity information than pitches.
For example, the $\text{P}_\text{ASI}$ of original BN features is 3.2\%, meaning that the adversary can recognize the identity of speakers from original BN features with 96.8\% accuracy  (recall that there are 921 different speakers).
In contrast, the $\text{P}_\text{ASI}$ of original pitch is $62\%$.

The empirical privacy of BN features and pitch improves significantly when using our DP extractors. For instance, using an analytical privacy budget of $\varepsilon=1$ improves the empirical privacy $\text{P}_\text{ASI}$ from the baseline of $3.2\%$ to above $86\%$ for BN features, and from $62\%$ to above $94\%$ for pitch. Figure~\ref{fig:DirectBN} also confirms that a better analytical privacy bound results in better empirical privacy. For example, the empirical privacy of DP BN features with $\varepsilon=100$ is 60\%, and it increases to 80\% for $\varepsilon=10$. 
We emphasize that although the reported analytical budget $\varepsilon$ for BN features is at the frame level, and thus translates (through the composition property of DP) into a large $\varepsilon'$ for the utterance level, the resulting noise appears to be sufficient in practice to protect well against re-identification attacks that exploit the entire sequence of BN features. We attribute this to the worst-case nature of DP (which often leads to large gaps between analytical and empirical privacy \cite{DP_eval,jagielski2020auditing,nasr2021adversary}) but also to the fact that our extractors are explicitly trained to preserve only the relevant linguistic and prosodic information.

\begin{figure}[t!]
\pgfplotsset{
    every non boxed x axis/.style={},
    boxplot/every box/.style={solid,ultra thin,black},
    boxplot/every whisker/.style={solid,ultra thin,black},
    boxplot/every median/.style={solid,very thick, red},
}
\pgfplotstableread{tikz/DirectBN/WER.txt}\wer
\pgfplotstableread{tikz/DirectBN/WER_NaiveBaseline.txt}\werNaive
\pgfplotstableread{tikz/DirectPitch/Corr.txt}\corr
\pgfplotstableread{tikz/DirectPitch/Corr_NaiveBaselien.txt}\corrNaive
\centering
\begin{tikzpicture}
\node[inner sep=0pt] (whitehead) at (0.5,-0.2)
{\footnotesize Baseline};
\node[inner sep=0pt] (whitehead) at (1.95,-0.2)
{\footnotesize DP BN};
    \footnotesize
        \begin{axis}[
        axis x line=bottom,
        axis y line=left,
        width=4.75cm,
        height=3.5cm,
        boxplot/draw direction=y,
        xtick={1,2.5,3.5,4.5,5.5},
        ymin=0,ymax=100,
title=BN features,
xticklabels={\vphantom{.1cm}$\varepsilon$=$\infty$,$\varepsilon$=100,~$\varepsilon$=10,$\varepsilon$=1,$\varepsilon$=0.5},
        xticklabel style={yshift=-10pt},
        ylabel={$\text{U}_\text{ASR}$ (\%)},
        y label style={at={(axis description cs:-.15,.5)},anchor=south},
        xtick style={draw=none},
        ytick style={draw=none},
]
        \addplot+[boxplot, boxplot/draw position=1,mark=*, mark options={black,scale=0},boxplot/box extend=0.5,boxplot/every median/.style={solid,very thick, black}] table[y=einf]{\wer};
\addplot+[boxplot, boxplot/draw position=2.5,mark=*, mark options={red,scale=0},boxplot/box extend=0.5] table[y=e100]{\wer};
\addplot+[boxplot, boxplot/draw position=3.5,mark=*, mark options={red,scale=0},boxplot/box extend=0.5] table[y=e10]{\wer};
\addplot+[boxplot, boxplot/draw position=4.5,mark=*, mark options={red,scale=0},boxplot/box extend=0.5] table[y=e1]{\wer};
\addplot+[boxplot, boxplot/draw position=5.5,mark=*, mark options={red,scale=0},boxplot/box extend=0.5] table[y=e05]{\wer};
\addplot+[boxplot, boxplot/draw position=2.5,mark=*, mark options={blue,scale=0},boxplot/box extend=0.5,boxplot/every median/.style={solid,very thick, blue}] table[y=e100]{\werNaive};
\addplot+[boxplot, boxplot/draw position=3.5,mark=*, mark options={blue,scale=0},boxplot/box extend=0.5,boxplot/every median/.style={solid,very thick, blue}] table[y=e10]{\werNaive};
\addplot+[boxplot, boxplot/draw position=4.5,mark=*, mark options={blue,scale=0},boxplot/box extend=0.5,boxplot/every median/.style={solid,very thick, blue}] table[y=e1]{\werNaive};
\addplot+[boxplot, boxplot/draw position=5.5,mark=*, mark options={blue,scale=0},boxplot/box extend=0.5,boxplot/every median/.style={solid,very thick, blue}] table[y=e1]{\werNaive};
\end{axis}
        \end{tikzpicture}
        \hspace{.2cm}
        \begin{tikzpicture}
\node[inner sep=0pt] (whitehead) at (0.5,-0.2)
{\footnotesize Baseline};
\node[inner sep=0pt] (whitehead) at (1.95,-0.2)
{\footnotesize DP Pitch};
    \footnotesize
        \begin{axis}[
        axis x line=bottom,
        axis y line=left,
        ymin=-0.2,ymax=1,
        width=4.75cm,
        height=3.5cm,
        boxplot/draw direction=y,
        xtick={1,2.5,3.5,4.5,5.5},
ytick={-0.2,0,0.2,0.4,0.6,0.8,1.0},
        title=Pitch,
        yticklabels={-0.2,0.0,0.2,0.4,0.6,0.8,1.0},
        xticklabels={\vphantom{.1cm}$\varepsilon$=$\infty$,$\varepsilon$=100,~$\varepsilon$=10,$\varepsilon$=5,$\varepsilon$=1},
        xticklabel style={yshift=-10pt},
        ylabel={Correlation},
        y label style={at={(axis description cs:-.15,.5)},anchor=south},
        xtick style={draw=none},
        ytick style={draw=none},
]
\addplot+[boxplot, boxplot/draw position=1,mark=*, mark options={scale=0.5},boxplot/box extend=0.5,boxplot/every median/.style={solid,very thick, black}] table[y=einf]{\corr};
\addplot+[boxplot, boxplot/draw position=2.5,mark=*, mark options={red,scale=0},boxplot/box extend=0.5] table[y=e100]{\corr};
        \addplot+[boxplot, boxplot/draw position=2.5,mark=*, mark options={blue,scale=0},boxplot/box extend=0.5,boxplot/every median/.style={solid,very thick, blue}] table[y index=2]{\corrNaive};
\addplot+[boxplot, boxplot/draw position=3.5,mark=*, mark options={red,scale=0},boxplot/box extend=0.5] table[y=e10]{\corr};
        \addplot+[boxplot, boxplot/draw position=3.5,mark=*, mark options={blue,scale=0},boxplot/box extend=0.5,boxplot/every median/.style={solid,very thick, blue}] table[y index=1]{\corrNaive};
\addplot+[boxplot, boxplot/draw position=4.5,boxplot/box extend=0.5] table[y=e5]{\corr};
         \addplot+[boxplot,boxplot/draw position=4.5,boxplot/box extend=0.5,boxplot/every median/.style={solid,very thick, blue}] table[y index=2]{\corrNaive};
         \addplot+[boxplot, boxplot/draw position=5.5,boxplot/box extend=0.5] table[y=e1]{\corr};
         \addplot+[boxplot,boxplot/draw position=5.5,boxplot/box extend=0.5,boxplot/every median/.style={solid,very thick, blue}] table[y index=0]{\corrNaive};

\end{axis}
        \end{tikzpicture}
    \caption{Utility of \textcolor{black}{original BN features} versus our \textcolor{blue}{naive DP BN baseline} and \textcolor{red}{our proposed DP BN features} (left), and \textcolor{black}{original pitch} versus our \textcolor{blue}{naive DP pitch baseline} and \textcolor{red}{proposed DP pitch} (right) for different privacy budgets $\varepsilon$. The utility of BN features is assessed by the performance $\text{U}_\text{ASR}$ of an ASR system trained using the corresponding BN features. The utility of pitch is assessed by its correlation to the original pitch. The impact of our DP extractors on utility is negligible for BN features. It is more marked for pitch as $\epsilon$ gets small. Yet, utility is much better than with the naive DP baseline.}
\label{fig:DirectBNUtility}
\end{figure}
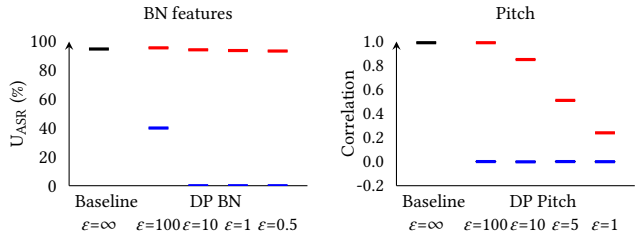

\looseness=-1 \myparagraph{How does our approach affect the utility of  pitch and BN features?}
Figure~\ref{fig:DirectBNUtility} shows the impact of using our DP extractors on the utility of the resulting pitch and BN features. The utility of BN features is measured by the performance of an ASR model trained from the BN features, while the utility of pitch is measured by its correlation to the original pitch sequence. In addition to pitch and BN features obtained with our DP extractors, we also report the utility of naive DP baselines based on directly adding noise to the original features (see Appendix~\ref{app:dp_baselines} for details). Unsurprisingly, these baselines perform very poorly, essentially destroying the utility even for large values of $\varepsilon$. In contrast, our DP BN features induce a negligible drop in the performance of the ASR model. Therefore, our DP BN extractor outputs high-utility BN features that contain enough linguistic information to perform the transcription task, while effectively concealing speaker information as shown in Figure~\ref{fig:DirectBN}. 

Regarding pitch, Figure~\ref{fig:DirectBNUtility} shows that the more noise injected in the pitch, the less correlation with the original pitch. While this is expected, we see that our approach is able to maintain high correlation with the original pitch sequence as long as $\varepsilon$ is not too small (in contrast to the naive baseline).

\subsection{Privacy and Utility of Anonymized Speech}
\label{sec:SeperatePitchBN}

We now plug the pitch and BN extractors into the full speaker anonymization pipeline (Figure~\ref{fig:BlockDiagram}) to generate anonymized utterances, and empirically evaluate their privacy and utility. Note that for conciseness, we focus on $\varepsilon \in \{1,10,100\}$ in the following.

\myparagraph{Separate effect of DP pitch and DP BN.}
We start by evaluating the effect of DP pitch and DP BN separately.
Figure~\ref{fig:DpPitchVPC} shows the empirical privacy and utility of \AnonDPPitch for different $\varepsilon$ against the state-of-the-art approach {\AnonPC}. Decreasing $\varepsilon$ in the DP pitch extractor improves the protection of utterances anonymized with \AnonDPPitch against speaker linkage attacks, as reflected by both empirical privacy metrics $\text{P}_\text{ASV,e}$ and $ \text{P}_\text{ASV,l}$. On the other hand, the impact on utility, as measured by the performance $\text{U}_\text{ASR}$ of the ASR model, is quite negligible.

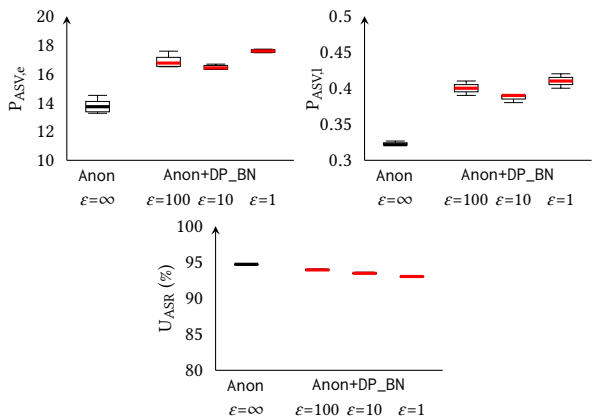
\begin{figure}[t!]
\pgfplotsset{
    every non boxed x axis/.style={},
    boxplot/every box/.style={solid,ultra thin,black},
    boxplot/every whisker/.style={solid,ultra thin,black},
    boxplot/every median/.style={solid,very thick, red},
}
\centering
\pgfplotstableread{PDRR/DPBN/PDRR_DPBN.txt}\DPBN
\begin{tikzpicture}
\node[inner sep=0pt] (whitehead) at (0.4,-0.2)
{\footnotesize \Anon};
\node[inner sep=0pt] (whitehead) at (1.9,-0.2)
{\footnotesize \AnonDPBN};
    \footnotesize
        \begin{axis}[
        axis x line=bottom,
        axis y line=left,
        width=4.6cm,
        height=3.5cm,
        boxplot/draw direction=y,
        xtick={1,2.5,3.5,4.5},
        ymin=10,ymax=20,
xticklabels={\vphantom{.1cm}$\varepsilon$=$\infty$,$\varepsilon$=100,$\varepsilon$=10,$\varepsilon$=1},
        xticklabel style={yshift=-10pt},
        ylabel={$ \text{P}_\text{ASV,e}$},
        y label style={at={(axis description cs:-.15,.5)},anchor=south},
        xtick style={draw=none},
        ytick style={draw=none},
]
        \addplot+[boxplot, boxplot/draw position=1,mark=*, mark options={black,scale=0},boxplot/box extend=0.5,boxplot/box extend=0.5,boxplot/box extend=0.5,boxplot/every median/.style={solid,very thick, black}] table[y=einfE]{\DPBN};
\addplot+[boxplot, boxplot/draw position=2.5,mark=*, mark options={red,scale=0},boxplot/box extend=0.5] table[y=e100E]{\DPBN};
\addplot+[boxplot, boxplot/draw position=3.5,mark=*, mark options={red,scale=0},boxplot/box extend=0.5] table[y=e10E]{\DPBN};
\addplot+[boxplot, boxplot/draw position=4.5,mark=*, mark options={red,scale=0},boxplot/box extend=0.5] table[y=e1E]{\DPBN};
\end{axis}
        \end{tikzpicture}    
\begin{tikzpicture}
\node[inner sep=0pt] (whitehead) at (0.4,-0.2)
{\footnotesize \Anon};
\node[inner sep=0pt] (whitehead) at (1.9,-0.2)
{\footnotesize \AnonDPBN};
    \footnotesize
        \begin{axis}[
        axis x line=bottom,
        axis y line=left,
        width=4.6cm,
        height=3.5cm,
        boxplot/draw direction=y,
        xtick={1,2.5,3.5,4.5},
        ymin=0.3,ymax=0.5,
xticklabels={\vphantom{.1cm}$\varepsilon$=$\infty$,$\varepsilon$=100,$\varepsilon$=10,$\varepsilon$=1},
        xticklabel style={yshift=-10pt},
        ylabel={$ \text{P}_\text{ASV,l}$},
        y label style={at={(axis description cs:-.15,.5)},anchor=south},
        xtick style={draw=none},
        ytick style={draw=none},
]
        \addplot+[boxplot, boxplot/draw position=1,mark=*, mark options={black,scale=0},boxplot/box extend=0.5,boxplot/box extend=0.5,boxplot/box extend=0.5,boxplot/every median/.style={solid,very thick, black}] table[y=einfU]{\DPBN};
\addplot+[boxplot, boxplot/draw position=2.5,mark=*, mark options={red,scale=0},boxplot/box extend=0.5] table[y=e100U]{\DPBN};
\addplot+[boxplot, boxplot/draw position=3.5,mark=*, mark options={red,scale=0},boxplot/box extend=0.5] table[y=e10U]{\DPBN};
\addplot+[boxplot, boxplot/draw position=4.5,mark=*, mark options={red,scale=0},boxplot/box extend=0.5] table[y=e1U]{\DPBN};

        \end{axis}
        \end{tikzpicture}

    \begin{tikzpicture}
\node[inner sep=0pt] (whitehead) at (0.4,-0.2)
{\footnotesize \Anon};
\node[inner sep=0pt] (whitehead) at (1.9,-0.2)
{\footnotesize \AnonDPBN};
    \footnotesize
        \begin{axis}[
        axis x line=bottom,
        axis y line=left,
        width=4.6cm,
        height=3.5cm,
        boxplot/draw direction=y,
        xtick={1,2.5,3.5,4.5},
        ymin=80,ymax=100,
xticklabels={\vphantom{.1cm}$\varepsilon$=$\infty$,$\varepsilon$=100,$\varepsilon$=10,$\varepsilon$=1},
        xticklabel style={yshift=-10pt},
        ylabel={$ \text{U}_\text{ASR}$ (\%)},
        y label style={at={(axis description cs:-.15,.5)},anchor=south},
        xtick style={draw=none},
        ytick style={draw=none},
]
        \addplot+[boxplot, boxplot/draw position=1,mark=*, mark options={black,scale=0},boxplot/box extend=0.5,boxplot/box extend=0.5,boxplot/box extend=0.5,boxplot/every median/.style={solid,very thick, black}] table[y=einfW]{\DPBN};
\addplot+[boxplot, boxplot/draw position=2.5,mark=*, mark options={red,scale=0},boxplot/box extend=0.5] table[y=e100W]{\DPBN};
\addplot+[boxplot, boxplot/draw position=3.5,mark=*, mark options={red,scale=0},boxplot/box extend=0.5] table[y=e10W]{\DPBN};
\addplot+[boxplot, boxplot/draw position=4.5, mark=*, mark options={red,scale=0},boxplot/box extend=0.5] table[y=e1W]{\DPBN};

        \end{axis}
        \end{tikzpicture}       
    \caption{Empirical privacy (top) and utility (bottom) of utterances anonymized with {\Anon} and our proposed \textcolor{red}{\AnonDPBN} for different privacy budgets $\varepsilon$. Empirical privacy is measured by the EER ($ \text{P}_\text{ASV,e}$) and unlinkability ($ \text{P}_\text{ASV,l}$) of a speaker linkage attack, while utility is assessed by the performance $\text{U}_\text{ASR}$ of an ASR system trained on anonymized utterances. Boxplots are computed over 5 runs of x-vector selection. Plugging our DP BN extractor in the state-of-the-art speaker anonymization pipeline improves privacy at a negligible loss in utility.}
\label{fig:DpBNVPC}
\end{figure}

Figure~\ref{fig:DpBNVPC} compares the empirical privacy and utility of \AnonDPBN for different $\varepsilon$ against \Anon. Again, we see that utterances anonymized by \AnonDPBN are better protected against speaker linkage attacks. The loss in utility (as measured by $\text{U}_\text{ASR}$) is slightly more noticeable than for \AnonDPPitch, which is expected since BN features contain most of the linguistic information. Nevertheless, the utility remains high: \Anon achieves utility $94.69\%$, while \AnonDPBN achieves $93.96\%$, $93.49\%$ and $93.03\%$ for privacy budgets of $100$, $10$ and $1$ respectively.

\begin{figure}[t!]
\pgfplotsset{
    every non boxed x axis/.style={},
    boxplot/every box/.style={solid,ultra thin,black},
    boxplot/every whisker/.style={solid,ultra thin,black},
    boxplot/every median/.style={solid,very thick, red},
}
\centering
\pgfplotstableread{PDRR/DPpitch/PDRR_DPpitch.txt}\DPPitch
\begin{tikzpicture}
\node[inner sep=0pt] (whitehead) at (0.5,-0.2)
{\footnotesize \AnonPC};
\node[inner sep=0pt] (whitehead) at (1.9,-0.2)
{\footnotesize \AnonDPPitch};
    \footnotesize
        \begin{axis}[
        axis x line=bottom,
        axis y line=left,
        width=4.6cm,
        height=3.5cm,
        boxplot/draw direction=y,
        xtick={1,2.5,3.5,4.5},
        ymin=10,ymax=20,
xticklabels={\vphantom{.1cm}$\varepsilon$=$\infty$,$\varepsilon$=100,$\varepsilon$=10,$\varepsilon$=1},
        xticklabel style={yshift=-10pt},
        ylabel={$ \text{P}_\text{ASV,e}$},
        y label style={at={(axis description cs:-.15,.5)},anchor=south},
        xtick style={draw=none},
        ytick style={draw=none},
]
        \addplot+[boxplot, boxplot/draw position=1,mark=*, mark options={black,scale=0},boxplot/box extend=0.5,boxplot/box extend=0.5,boxplot/every median/.style={solid,very thick, black}] table[y=einfE]{\DPPitch};
\addplot+[boxplot, boxplot/draw position=2.5,mark=*, mark options={red,scale=0},boxplot/box extend=0.5] table[y=e100E]{\DPPitch};
\addplot+[boxplot, boxplot/draw position=3.5,mark=*, mark options={red,scale=0},boxplot/box extend=0.5] table[y=e10E]{\DPPitch};
\addplot+[boxplot, boxplot/draw position=4.5,mark=*, mark options={red,scale=0},boxplot/box extend=0.5] table[y=e1E]{\DPPitch};
\end{axis}
        \end{tikzpicture}    
\begin{tikzpicture}
\node[inner sep=0pt] (whitehead) at (0.5,-0.2)
{\footnotesize \AnonPC};
\node[inner sep=0pt] (whitehead) at (1.9,-0.2)
{\footnotesize \AnonDPPitch};
    \footnotesize
        \begin{axis}[
        axis x line=bottom,
        axis y line=left,
        width=4.6cm,
        height=3.5cm,
        boxplot/draw direction=y,
        xtick={1,2.5,3.5,4.5},
        ymin=0.3,ymax=0.5,
xticklabels={\vphantom{.1cm}$\varepsilon$=$\infty$,$\varepsilon$=100,$\varepsilon$=10,$\varepsilon$=1},
        xticklabel style={yshift=-10pt},
        ylabel={$ \text{P}_\text{ASV,l}$},
        y label style={at={(axis description cs:-.15,.5)},anchor=south},
        xtick style={draw=none},
        ytick style={draw=none},
]
        \addplot+[boxplot, boxplot/draw position=1,mark=*, mark options={black,scale=0},boxplot/box extend=0.5,boxplot/box extend=0.5,boxplot/every median/.style={solid,very thick, black}] table[y=einfU]{\DPPitch};
\addplot+[boxplot, boxplot/draw position=2.5,mark=*, mark options={red,scale=0},boxplot/box extend=0.5] table[y=e100U]{\DPPitch};
\addplot+[boxplot, boxplot/draw position=3.5,mark=*, mark options={red,scale=0},boxplot/box extend=0.5] table[y=e10U]{\DPPitch};
\addplot+[boxplot, boxplot/draw position=4.5,mark=*, mark options={red,scale=0},boxplot/box extend=0.5] table[y=e1U]{\DPPitch};

        \end{axis}
        \end{tikzpicture}\\
        \begin{tikzpicture}
        \node[inner sep=0pt] (whitehead) at (0.5,-0.2)
{\footnotesize \AnonPC};
\node[inner sep=0pt] (whitehead) at (1.9,-0.2)
{\footnotesize \AnonDPPitch};
    \footnotesize
        \begin{axis}[
        axis x line=bottom,
        axis y line=left,
        width=4.6cm,
        height=3.5cm,
        boxplot/draw direction=y,
        xtick={1,2.5,3.5,4.5},
        ymin=80,ymax=100,
xticklabels={\vphantom{.1cm}$\varepsilon$=$\infty$,$\varepsilon$=100,$\varepsilon$=10,$\varepsilon$=1},
        xticklabel style={yshift=-10pt},
        ylabel={$ \text{U}_\text{ASR} \%$},
        y label style={at={(axis description cs:-.15,.5)},anchor=south},
        xtick style={draw=none},
        ytick style={draw=none},
]
        \addplot+[boxplot, boxplot/draw position=1,mark=*, mark options={black,scale=0},boxplot/box extend=0.5,boxplot/box extend=0.5,boxplot/every median/.style={solid,very thick, black}] table[y=einfW]{\DPPitch};
\addplot+[boxplot, boxplot/draw position=2.5,mark=*, mark options={red,scale=0},boxplot/box extend=0.5] table[y=e100W]{\DPPitch};
\addplot+[boxplot, boxplot/draw position=3.5,mark=*, mark options={red,scale=0},boxplot/box extend=0.5] table[y=e10W]{\DPPitch};
\addplot+[boxplot, boxplot/draw position=4.5,mark=*, mark options={red,scale=0},boxplot/box extend=0.5] table[y=e1W]{\DPPitch};
        \end{axis}
        \end{tikzpicture} 
        
    \caption{\looseness=-1 Empirical privacy (top) and utility (bottom) of utterances anonymized with \AnonPC and our proposed \textcolor{red}{\AnonDPPitch} for different privacy budgets $\varepsilon$. Empirical privacy is measured by the EER ($ \text{P}_\text{ASV,e}$) and unlinkability ($ \text{P}_\text{ASV,l}$) of a speaker linkage attack, while utility is assessed by the performance $\text{U}_\text{ASR}$ of an ASR system trained on anonymized utterances. Boxplots are computed over 5 runs of x-vector selection. Plugging our DP pitch extractor in the state-of-the-art speaker anonymization pipeline improves privacy with almost no impact on utility.}
\label{fig:DpPitchVPC}
\end{figure}

All figures report the variation of the results due to the randomness in x-vector selection. In Appendix~\ref{sec:NoiseEffect}, we report the variation due to the randomness of the noise, which we typically found to be of smaller magnitude.

\myparagraph{Evaluation of the complete system.}
Finally, we compare our full DP speaker anonymization approach with both DP pitch and DP BN features (\DP) to the state-of-the-art technique of the VoicePrivacy challenge (\AnonPC) \cite{srivastava:hal-03197376,tomashenko2020introducing}. Table~\ref{tab:AllTogether} reports the privacy and utility metrics achieved by our approach for different privacy budgets $\varepsilon$. In all cases, \DP provides significantly better protection against speaker linkage attacks at a negligible cost in utility. Indeed, both empirical privacy metrics $\text{P}_\text{ASV,e}$ and $ \text{P}_\text{ASV,l}$ nearly double for a very small loss in ASR performance (as measured by $\text{U}_\text{ASR}$). For example, our method \DP with $\varepsilon=100$ for both BN and pitch extractors achieves 24.22\% EER, 0.57 unlinkability and 94.00\% WER, whereas \AnonPC achieves 14.85\% EER, 0.35 unlinkability and 94.64\% WER. As desired, decreasing $\varepsilon$ (i.e., enforcing stronger analytical privacy guarantees) in \DP improves the empirical privacy. For example, \DP with $\varepsilon=1$ increases the empirical privacy of \DP with $\varepsilon=10$ from $26.68\%$ and $0.65$ unlinkability to $29.98\%$ and $0.70$, with less than $1\%$ degradation in WER.
Comparing the results in Table~\ref{tab:AllTogether} with those in Figures~\ref{fig:DpPitchVPC}-\ref{fig:DpBNVPC} also shows that reducing speaker information in \emph{both} pitch and BN features provides a large gain in the privacy of anonymized utterances. These results fully validate our design choices and demonstrate the usefulness of our approach for speaker anonymization.

\begin{table}[t] 
    \centering 
    \footnotesize
\caption{Empirical privacy and utility of our complete speaker anonymization method {\DP} for different analytical privacy budgets $\varepsilon$ against the state-of-the-art approach {\AnonPC}. Results are computed across 5 runs of x-vector selection. The results show that our method allows to significantly improve the privacy of anonymized speech at a negligible cost in utility. } 
       \begin{tabular}{lccccc} 
    \toprule  
    \multirow{3}{*}{\textbf{Method}} & \multicolumn{4}{c}{\textbf{Privacy}} &  \multirow{1}{*}{\textbf{Utility}} \\ 
     & \multicolumn{2}{c}{Analytical ($\varepsilon$)} & \multicolumn{2}{c}{Empirical} &  Empirical \\ 
     & BN  &  Pitch  & $ \text{P}_\text{ASV,e}$ (\%) & $ \text{P}_\text{ASV,l}$ & $ \text{U}_\text{ASR}$ (\%) \\
    \hline
    \textbf{\AnonPC} & $\infty$ & $\infty$ & $14.62 \pm 0.25$ & $.35 \pm .01$ & $94.64 \pm .06$\\
    \textbf{\DP} & $100$  & $100$  & $24.22 \pm .44$ & $.57 \pm .01$ & $94.00 \pm .10 $\\
    \textbf{\DP} & $10$  & $10$  & $27.68 \pm .25$ & $.65 \pm .01$ & $93.01 \pm .07 $\\
    \textbf{\DP} & $1$  & $1$  & $29.98 \pm .76$ & $.70 \pm .01$ & $92.16 \pm .05 $\\

    \bottomrule
    \end{tabular}
\label{tab:AllTogether}
\end{table}

\begin{figure}[t!]
\pgfplotsset{
    every non boxed x axis/.style={},
    boxplot/every box/.style={solid,ultra thin,black},
    boxplot/every whisker/.style={solid,ultra thin,black},
    boxplot/every median/.style={solid,very thick, red},
}
\centering
\pgfplotstableread{DisparateImpact/disparate.txt}\Disparate
\begin{tikzpicture}
    \footnotesize
        \begin{axis}[
        axis x line=bottom,
        axis y line=left,
        width=4.6cm,
        height=3.5cm,
        boxplot/draw direction=y,
        xtick={1,2},
        ymin=0,ymax=50,
        ytick={0,10,...,50},
yticklabels={0,10,...,50},
        xticklabels={\AnonPC,\DP},
        ylabel={$ \text{P}_\text{ASV,e}$},
        y label style={at={(axis description cs:-0.15,.5)},anchor=south},
        xtick style={draw=none},
        ytick style={draw=none},
]
        \addplot+[boxplot, boxplot/draw position=1,mark=*, mark options={red,scale=0},boxplot/box extend=0.5] table[y expr=\thisrow{AnonPC_eer_mean}*100]{\Disparate};
        
        \addplot+[boxplot, boxplot/draw position=2,mark=*, mark options={red,scale=0},boxplot/box extend=0.5] table[y expr=\thisrow{AnonDP_10_10_eer_mean}*100]{\Disparate};
        \end{axis}
        \end{tikzpicture}
\begin{tikzpicture}
    \footnotesize
        \begin{axis}[
        axis x line=bottom,
        axis y line=left,
        width=4.6cm,
        height=3.5cm,
        boxplot/draw direction=y,
        xtick={1,2},
        ymin=50,ymax=100,
        ytick={50,60,...,100},
yticklabels={50,60,...,100},
        xticklabels={\AnonPC,\DP},
        ylabel={$ \text{U}_\text{ASR} \%$},
        y label style={at={(axis description cs:-0.15,.5)},anchor=south},
        xtick style={draw=none},
        ytick style={draw=none},
]
        \addplot+[boxplot, boxplot/draw position=1,mark=*, mark options={red,scale=0},boxplot/box extend=0.5] table[y expr=(1-\thisrow{AnonPC_wer_mean})*100]{\Disparate};
        
        \addplot+[boxplot, boxplot/draw position=2,mark=*, mark options={red,scale=0},boxplot/box extend=0.5] table[y expr=(1-\thisrow{AnonDP_10_10_wer_mean})*100]{\Disparate};
        \end{axis}
        \end{tikzpicture}

    \caption{Variation in the privacy and utility across speakers for the \AnonPC baseline and our \DP scheme. Boxplots are computed over the 29 speakers of the enrollment set.}
\label{fig:PerSpeakerDispareteImpact}
\end{figure}
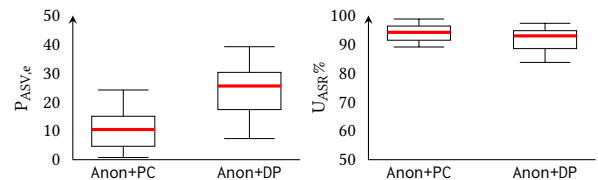

\looseness=-1 \myparagraph{Variation across speakers.}
We investigate how the privacy-utility trade-off varies across (subpopulations) of speakers. Note that LibriSpeech contains an imbalanced number of male/female speakers and each speaker has various number of utterances, which may cause some variations across sex and speaker.
Table~\ref{tab:PerGenderDisparateImpact} and Figure~\ref{fig:PerSpeakerDispareteImpact} compare the per-sex and per-speaker performance of our approach with the \AnonPC baseline. The results show that the privacy-utility trade-offs are similar across sex for both our method and the baseline.
We observe stronger variations of privacy and utility across speakers for both \AnonPC and \DP. Yet, the gains provided by our approach are clear: while the distribution of utility across speakers is similar for both methods, \DP ``shifts up'' the distribution of privacy: for instance, the worst privacy protection across speakers with \DP is roughly the same as the median privacy protection with \AnonPC. Similarly, \DP protects half of the speakers better than what \AnonPC provides for the best protected speaker.
Finally, while differential privacy can have a disparate impact on the utility across different subpopulations in certain settings \cite{BagdasaryanPS19}, we note that it does not seem to be the case for our approach compared to the baseline \AnonPC. Indeed, the magnitude of the variations (across sex and speaker) remain roughly the same across both methods.

\begin{table}[t] 
    \centering 
    \footnotesize
\caption{Privacy and utility of the \AnonPC baseline and our \DP scheme on female and male subpopulations.} 
       \begin{tabular}{lcccccc} 
    \toprule  
    \multirow{3}{*}{\textbf{Method}} & \multicolumn{4}{c}{\textbf{Privacy}} &  \multicolumn{2}{c}{\textbf{Utility}} \\ 
     & \multicolumn{2}{c}{Analytical ($\varepsilon$)} & \multicolumn{2}{c}{$\text{P}_\text{ASV,e}$} &  \multicolumn{2}{c}{$ \text{U}_\text{ASR}$} \\ 
     & BN  &  Pitch  & Female &  Male &  Female & Male \\
    \hline
    \textbf{\AnonPC} & $\infty$ & $\infty$  & $15.87$ & $13.09$ & $94.25$ & $94.24$\\
    \textbf{\DP} & $10$ & $10$ & $28.51$ & $26.34$ & $92.40$ & $92.56$\\

    \bottomrule
    \end{tabular}
\label{tab:PerGenderDisparateImpact}
\end{table}

\section{Discussion and Future Work}
\label{sec:concl}
In this paper, we proposed a differentially private speaker anonymization approach which can be used to share speech utterances for training or deployment of voice-based services while concealing the speaker's identity. 
More specifically, we revisited the disentanglement of speaker (x-vector), linguistic (BN features) and prosodic (pitch) information used in state-of-the-art speaker anonymization techniques so as to analytically bound the speaker information contained in pitch and BN features using carefully designed extractors and differential privacy. Plugging our proposed DP pitch and BN extractors in a speaker anonymization pipeline, we are able to re-synthesize utterances in a differentially private fashion. We also empirically demonstrated that our approach provides significant gains in practical privacy protection against strong attacks while maintaining a high level of utility. 

Below, we discuss some limitations of our approach and promising directions for future work.

\myparagraph{Tightness of analytical privacy guarantees.} In our empirical study, we observed that our approach provides high protection against concrete attacks despite the looseness of the analytical DP guarantee at the utterance-level. Recall that the $\varepsilon$ we report for BN features is frame-level and should be multiplied by the utterance length to obtain an utterance-level guarantee. Our analytical guarantees are even weaker if we consider the DP guarantee at the user level (i.e., adding the budget of all utterances contributed by a given user). While this gap between empirical and analytical privacy is expected and commonly observed in DP literature \cite{DP_eval,jagielski2020auditing,nasr2021adversary}, we briefly discuss below several ways in which it could be reduced.

An obvious solution to get stronger analytical utterance-level privacy guarantees is simply to train DP BN features with a smaller $\varepsilon$. However, we could not manage to make the training converge to a satisfactory value with $\varepsilon<0.5$ (e.g., for $\varepsilon=0.1$), presumably because the signal to noise ratio becomes too small.
As an alternative, recent tools from DP theory may be leveraged to bound the analytical privacy budget more tightly for utterance and user-level DP. Instead of using simple composition which makes the utterance-level budget $\varepsilon'$ grow linearly with the utterance length $K$, one can resort to advanced composition \cite{Dwork2014a,kairouz2015composition} to achieve $\varepsilon'$ of order $\sqrt{K}\cdot \varepsilon$ at the cost of achieving slightly weaker $(\varepsilon, \delta)$-DP (we illustrate this in Appendix~\ref{app:tight}). One may also reduce the length of utterances by slicing them into shorter segments, which was recently shown to preserve utility in x-vector based speaker anonymization \cite{slicing}. When the anonymized corpus is large,
since the adversary does not know which user submitted which utterance, we can further benefit from privacy amplification by shuffling \cite{erlingsson2019amplification,balle2019privacy,clones}, which gives a reduction of the privacy budget of order $1/\sqrt{N}$ where $N$ is the total number of utterances shared by all users. 

Despite these potential improvements, we stress that there is a fundamental gap between the notion of DP we enforce and what we are truly after (we want to remove information about speaker identity, whereas DP seeks to make any two utterances sufficiently indistinguishable). We believe that the careful design of appropriate relaxations of DP that would better capture the objective of speaker anonymization constitutes an interesting challenge.

\looseness=-1 \myparagraph{Utility measures.} Another interesting future direction is to consider better utility measures for anonymized utterances and pitch. In this work, we demonstrated the utility of utterances through ASR performance.
While listening to a few samples suggests that the anonymized speech is intelligible to humans, this could be confirmed and quantified through subjective evaluations. Regarding pitch, we used the correlation with the original pitch as a proxy to measure the preservation of prosodic content.
Nonetheless, further experimentation is required to better quantify utility, for instance by training a network to predict certain prosodic attributes from pitch.

\myparagraph{Speech rate.} Our anonymization scheme preserves the speech rate as this is a requirement of the VoicePrivacy challenge \cite{tomashenko2020introducing}. However, a small amount of speaker information might be leaked through the speech rate. An interesting future direction is to address this potential privacy leakage without harming the utility of utterances. This is a challenging problem because the duration of some phonemes must remain unchanged to preserve utility and prosody, while the duration of silences and other phonemes may be changed more or less depending on their position inside words and the utterance.

\begin{acks}
This work was supported in part by the French National Research Agency under project DEEP-PRIVACY (ANR-18-CE23-0018) and by the European Union’s Horizon 2020 Research and Innovation Program under Grant Agreements No. 825081 COMPRISE and No. 952215 TAILOR. Experiments presented in this paper were partially carried out using the Grid'5000 testbed, supported by a scientific interest group hosted by Inria and including CNRS, RENATER and several Universities as well as other organizations (see \url{https://www.grid5000.fr}). Ali Shahin Shamsabadi and Nicolas Papernot were also partially supported by CIFAR and the DARPA GARD program. Finally, Ali Shahin Shamsabadi acknowledges the partial support from The Alan Turing Institute.
\end{acks}

\bibliographystyle{plain}
\bibliography{main}

%% file: appendix.tex
\section{Details on Experimental Setup}
\label{app:details_setup}

In this section, we give additional details on our experimental setup and implementations.

\subsection{Details on the Data Set}
\label{app:data_set}
The five subsets of LibriSpeech data set \cite{panayotov2015librispeech}, and the subset of LibriTTS \cite{libritts} used in our experiments, are detailed in Table~\ref{tab:LibriSpeech}. Note that we do not use the dev\_clean subset.

\subsection{Details on Speaker Anonymization Systems}
\label{app:anon_systems}

\myparagraph{DP pitch autoencoder.} We implement our DP pitch autoencoder in PyTorch and train it on train\_clean\_100, using a mini-batch size of 1 due to the variable sequence length. We use the Adam optimizer \cite{adam} with a learning rate of $1\text{e}^{-3}$, a weight decay of $1\text{e}^{-5}$, and a dropout of $1\text{e}^{-3}$, similarly to \cite{shamsabadi2021foolhd}.

\myparagraph{Target x-vector selection.}
In all speaker anonymization systems, the target x-vector for each utterance is selected as follows. First, we cluster all public x-vectors using the Affinity Propagation algorithm \cite{dueck2009affinity} and PLDA \cite{kenny2010} (obtaining 80 clusters in total). Second, we randomly select one cluster from the 10 largest clusters. Third, we randomly select half of the members of the dense cluster to introduce further randomness in the choice of x-vector.\footnote{We noticed that selecting less than 50\% of a cluster negatively affects utility.} Finally, we average the selected candidate x-vectors to obtain the target x-vector. This selection strategy is very similar to the ``dense'' strategy proposed in \cite{srivastava2020design}, but makes the choice of x-vector completely independent from the input utterance so that this step does not leak any information about the source speaker.

\subsection{Additional Details on Attacks}
\label{app:attacks}

We give additional details on the implementations of the ASI and ASV systems which form the basis of our attacks.

\myparagraph{ASI system.} The ASI system follows the classical TDNN speaker classification architecture in Kaldi \cite{snyder2018x}: it is composed of 5 TDNN layers after the input layer, followed by a statistical pooling layer which computes the mean and standard deviation over all the frames to get the utterance-level context. This layer is followed by 2 TDNN layers and finally a softmax output layer.
This system is trained on pitch or BN features, with or without $\varepsilon$-DP depending on the system under attack. In the case of pitch, silent regions are removed using energy-based voice activity detection before training the ASI. We train the ASI system over the training split with 15 epochs using a mini-batch size of 64.

\looseness=-1 \myparagraph{ASV system.}
The ASV system, i.e., both the x-vector extractor and PLDA, follows the classical ASV recipe in Kaldi \cite{povey2011kaldi}.
The x-vector extractor (a TDNN with MFCCs as inputs) and PLDA are trained
over the train\_clean\_360 data set, anonymized using exactly the same method and parameters as used for anonymizing the utterances that the adversary wants to attack.
The target x-vector selection used by the attack is the same as the one used by the speaker anonymization system (see Appendix~\ref{app:anon_systems}).
The ASV system is then used to compute PLDA scores between trial and an enrollment utterances using x-vectors extracted from these utterances.

\subsection{Details on the ASR Evaluation Model}
\label{app:asr}

\looseness=-1 The training procedure and architecture of the ASR system used to compute the utility metric $\text{U}_\text{ASR}$ is similar to the one used to extract BN features in Section~\ref{para:asr-training}, except that we do not use any noise layer after the BN extractor: the BN features extracted from anonymized utterances are directly fed to the triphone classifier to compute the loss $\mathcal{L}_\text{ASR}$.
We train a different ASR system for each speaker anonymization scheme (e.g., for each value of $\varepsilon$) so that the ASR model can adapt to each scheme. During decoding, we use a large trigram language model $P(W)$ available at the openslr website.\footnote{\url{http://openslr.org/11/}}

\subsection{Naive DP Baselines}
\label{app:dp_baselines}

We explain below how the naive input perturbation DP baselines evaluated in Figure~\ref{fig:DirectBNUtility} are implemented.

\looseness=-1 \myparagraph{Naive DP pitch baseline.} This baseline consists in adding DP noise directly to the normalized pitches (zero mean and unit std). As pitch values are unbounded, we artificially bound them by clipping in a range. We select the range $[-4,+4]$, which includes most values while being sufficiently narrow to keep the sensitivity small.

\myparagraph{Naive DP BN baseline.} This baseline consists in applying the noise layer $\noise_B$ (which normalizes the features, adds noise and normalizes again, see Eq.~\ref{eq:perturbed-BN}) to the original BN features used by the state-of-the-art speaker anonymization method \Anon and re-training the triphone classifier on top of these fixed noisy features. Note that, unlike our proposed approach, this naive baseline does not re-train the BN extractor $\B$. Fine-tuning the original triphone classifier or re-training it from scratch yields similar results.
\begin{table}[t]
\footnotesize
  \caption{Statistics of the different subsets of the LibriSpeech (top row) and LibriTTS (bottom row) data sets.}
  \label{tab:LibriSpeech}
  \centering
  \begin{tabular}{ccccccc}
  \toprule
   \multirow{2}{*}{} & \multirow{2}{*}{\textbf{Subset}} &   \multirow{2}{*}{\textbf{Size (hrs.)}} & \multicolumn{3}{c}{\textbf{\#speakers}} &  \multirow{2}{*}{\textbf{\#utterances}} \\
   & &  & \textbf{Female} & \textbf{Male} & \textbf{Total} & \\ 
   \midrule
  \multirow{5}{*}{{\rotatebox[origin=c]{90}{Librispeech}}} & train-clean\_100 & 100.6 & 125 & 126   &  251 & 28,539 \\
  & train-clean\_360 & 363.6 & 439 & 482   &  921 & 104,014 \\
  & train-other\_500 & 496.7 & 564 & 602   &  1,166 & 148,688 \\
  & dev\_clean       &   5.4 &  20 &  20   &   40 & 2,703 \\
  & test\_clean      &   5.4 &  20 &  20   &   40 &  2,620 \\
  \midrule
  {{\rotatebox[origin=c]{90}{LibriTTS}}} & train-clean\_100 & 54 & 123 & 124 & 247 & 33,236\\
  \bottomrule
  \end{tabular}
\end{table} 
\begin{table}[t] 
\footnotesize
    \centering 
    \renewcommand{\arraystretch}{1.1}
    \caption{Effect of x-vector selection strategy used in anonymization and in the attack on the empirical privacy of speech anonymized with the state-of-the-art {\AnonPC}. Results are computed across 5 runs of x-vector selection.} 
       \begin{tabular}{llcc} 
    \toprule 
    \multicolumn{2}{c}{\textbf{X-vector Selection Strategy}}  & \multicolumn{2}{c}{\textbf{Empirical Privacy}}  \\
    Anonymization & Attack & $ \text{P}_\text{ASV,e}$ (\%) & $ \text{P}_\text{ASV,l}$  \\
    \midrule 
    \multirow{2}{*}{speaker-level} &  speaker-level &  $45.90 \pm 1.86$	& $0.86 \pm 0.04$ \\
     & utterance-level  &  $15.58 \pm 0.23$ &	$0.37 \pm 0.01$ \\
     \midrule
    \multirow{2}{*}{utterance-level} &  speaker-level &   $44.32 \pm 0.67$ &	$0.91 \pm 0.01$ \\
     &  utterance-level  & $14.62 \pm 0.25$ & $0.35 \pm 0.01$ \\

    \bottomrule
    \end{tabular}
    \label{tab:PDRSvsPDRR}
\end{table}
\begin{figure*}
\begin{minipage}[t]{8.5cm}
\pgfplotsset{
    every non boxed x axis/.style={},
    boxplot/every box/.style={solid,ultra thin,black},
    boxplot/every whisker/.style={solid,ultra thin,black},
    boxplot/every median/.style={solid,very thick, red},
}
\centering
\pgfplotstableread{PDRR/DPpitch/DPpitch_noiseRandomness.txt}\DPPitchRandomNoise
\begin{tikzpicture}
    \footnotesize
        \begin{axis}[
        axis x line=bottom,
        axis y line=left,
        width=4.6cm,
        height=3.5cm,
        boxplot/draw direction=y,
        xtick={1,2,3},
        ymin=10,ymax=20,
        xticklabels={100,10,1},
        ylabel={$ \text{P}_\text{ASV,e}$},
        y label style={at={(axis description cs:-0.15,.5)},anchor=south},
        xtick style={draw=none},
        ytick style={draw=none},
        xlabel={\footnotesize Privacy budget $\varepsilon$}
        ]
        \addplot+[boxplot, boxplot/draw position=1,mark=*, mark options={red,scale=0},boxplot/box extend=0.5] table[y=e100E]{\DPPitchRandomNoise};
        \addplot+[boxplot, boxplot/draw position=2,mark=*, mark options={red,scale=0},boxplot/box extend=0.5] table[y=e10E]{\DPPitchRandomNoise};
        \addplot+[boxplot, boxplot/draw position=3,mark=*, mark options={red,scale=0},boxplot/box extend=0.5] table[y=e1E]{\DPPitchRandomNoise};
        \end{axis}
        \end{tikzpicture}    
\begin{tikzpicture}
    \footnotesize
        \begin{axis}[
        axis x line=bottom,
        axis y line=left,
        width=4.6cm,
        height=3.5cm,
        boxplot/draw direction=y,
        xtick={1,2,3},
        ymin=0.3,ymax=0.5,
        xticklabels={100,10,1},
        ylabel={$ \text{P}_\text{ASV,l}$},
        y label style={at={(axis description cs:-0.15,.5)},anchor=south},
        xtick style={draw=none},
        ytick style={draw=none},
        xlabel={\footnotesize Privacy budget $\varepsilon$}
        ]
        %
        \addplot+[boxplot, boxplot/draw position=1,mark=*, mark options={red,scale=0},boxplot/box extend=0.5] table[y=e100U]{\DPPitchRandomNoise};
        \addplot+[boxplot, boxplot/draw position=2,mark=*, mark options={red,scale=0},boxplot/box extend=0.5] table[y=e10U]{\DPPitchRandomNoise};
        \addplot+[boxplot, boxplot/draw position=3,mark=*, mark options={red,scale=0},boxplot/box extend=0.5] table[y=e1U]{\DPPitchRandomNoise};
        \end{axis}
        \end{tikzpicture}\\
        \begin{tikzpicture}
    \footnotesize
        \begin{axis}[
        axis x line=bottom,
        axis y line=left,
        width=4.6cm,
        height=3.5cm,
        boxplot/draw direction=y,
        xtick={1,2,3},
        ymin=80,ymax=100,
        xticklabels={100,10,1},
        ylabel={$ \text{U}_\text{ASR} \%$},
        y label style={at={(axis description cs:-0.15,.5)},anchor=south},
        xtick style={draw=none},
        ytick style={draw=none},
        xlabel={\footnotesize Privacy budget $\varepsilon$}
        ]
        \addplot+[boxplot, boxplot/draw position=1,mark=*, mark options={red,scale=0},boxplot/box extend=0.5] table[y=e100W]{\DPPitchRandomNoise};
        \addplot+[boxplot, boxplot/draw position=2,mark=*, mark options={red,scale=0},boxplot/box extend=0.5] table[y=e10W]{\DPPitchRandomNoise};
        \addplot+[boxplot, boxplot/draw position=3,mark=*, mark options={red,scale=0},boxplot/box extend=0.5] table[y=e1W]{\DPPitchRandomNoise};
        \end{axis}
        \end{tikzpicture} 
    \caption{Empirical privacy (top) and utility (bottom) of utterances anonymized with our proposed \textcolor{red}{\AnonDPPitch} for different privacy budgets $\varepsilon$. Empirical privacy is measured by the EER ($ \text{P}_\text{ASV,e}$) and unlinkability ($ \text{P}_\text{ASV,l}$) of a speaker linkage attack, while utility is assessed by the performance $\text{U}_\text{ASR}$ of an ASR system trained on anonymized utterances. Unlike in Figure~\ref{fig:DpPitchVPC}, boxplots are computed over 5 runs of DP noise addition. The results show that the variations due to the randomness of noise in our DP pitch extractor are small.}
    \label{fig:DpPitchVPCRandomNoiseEffectFinal}
\end{minipage}
\hfill
\begin{minipage}[t]{8.5cm}
\pgfplotsset{
    every non boxed x axis/.style={},
    boxplot/every box/.style={solid,ultra thin,black},
    boxplot/every whisker/.style={solid,ultra thin,black},
    boxplot/every median/.style={solid,very thick, red},
}
\centering
\pgfplotstableread{PDRR/DPBN/DPBNRandomNoise.txt}\DPBNRandomNoise
\begin{tikzpicture}
    \footnotesize
        \begin{axis}[
        axis x line=bottom,
        axis y line=left,
        width=4.6cm,
        height=3.5cm,
        boxplot/draw direction=y,
        xtick={1,2,3},
        ymin=10,ymax=20,
        xticklabels={100,10,1},
        ylabel={$ \text{P}_\text{ASV,e}$},
        y label style={at={(axis description cs:-0.15,.5)},anchor=south},
        xtick style={draw=none},
        ytick style={draw=none},
        xlabel={\footnotesize Privacy budget $\varepsilon$}
        ]
        \addplot+[boxplot, boxplot/draw position=1,mark=*, mark options={red,scale=0},boxplot/box extend=0.5] table[y=e100E]{\DPBNRandomNoise};
        \addplot+[boxplot, boxplot/draw position=2,mark=*, mark options={red,scale=0},boxplot/box extend=0.5] table[y=e10E]{\DPBNRandomNoise};
        \addplot+[boxplot, boxplot/draw position=3,mark=*, mark options={red,scale=0},boxplot/box extend=0.5] table[y=e1E]{\DPBNRandomNoise};
        \end{axis}
        \end{tikzpicture}    
\begin{tikzpicture}
    \footnotesize
        \begin{axis}[
        axis x line=bottom,
        axis y line=left,
        width=4.6cm,
        height=3.5cm,
        boxplot/draw direction=y,
        xtick={1,2,3},
        ymin=0.3,ymax=0.5,
        xticklabels={100,10,1},
        ylabel={$ \text{P}_\text{ASV,l}$},
        y label style={at={(axis description cs:-0.15,.5)},anchor=south},
        xtick style={draw=none},
        ytick style={draw=none},
        xlabel={\footnotesize Privacy budget $\varepsilon$}
        ]
        %
        \addplot+[boxplot, boxplot/draw position=1,mark=*, mark options={red,scale=0},boxplot/box extend=0.5] table[y=e100U]{\DPBNRandomNoise};
        \addplot+[boxplot, boxplot/draw position=2,mark=*, mark options={red,scale=0},boxplot/box extend=0.5] table[y=e10U]{\DPBNRandomNoise};
        \addplot+[boxplot, boxplot/draw position=3,mark=*, mark options={red,scale=0},boxplot/box extend=0.5] table[y=e1U]{\DPBNRandomNoise};

        \end{axis}
        \end{tikzpicture}\\
        \begin{tikzpicture}
    \footnotesize
        \begin{axis}[
        axis x line=bottom,
        axis y line=left,
        width=4.6cm,
        height=3.5cm,
        boxplot/draw direction=y,
        xtick={1,2,3},
        ymin=80,ymax=100,
        xticklabels={100,10,1},
        ylabel={$ \text{U}_\text{ASR} \%$},
        y label style={at={(axis description cs:-0.15,.5)},anchor=south},
        xtick style={draw=none},
        ytick style={draw=none},
        xlabel={\footnotesize Privacy budget $\varepsilon$}
        ]
        \addplot+[boxplot, boxplot/draw position=1,mark=*, mark options={red,scale=0},boxplot/box extend=0.5] table[y=e100W]{\DPBNRandomNoise};
        \addplot+[boxplot, boxplot/draw position=2,mark=*, mark options={red,scale=0},boxplot/box extend=0.5] table[y=e10W]{\DPBNRandomNoise};
        \addplot+[boxplot, boxplot/draw position=3,mark=*, mark options={red,scale=0},boxplot/box extend=0.5] table[y=e1W]{\DPBNRandomNoise};

        \end{axis}
        \end{tikzpicture} 
        
    \caption{\looseness=-1 Empirical privacy (top) and utility (bottom) of utterances anonymized with our proposed \textcolor{red}{\AnonDPBN} for different privacy budgets $\varepsilon$. Empirical privacy is measured by the EER ($ \text{P}_\text{ASV,e}$) and unlinkability ($ \text{P}_\text{ASV,l}$) of a speaker linkage attack, while utility is assessed by the performance $\text{U}_\text{ASR}$ of an ASR system trained on anonymized utterances. Unlike in Figure~\ref{fig:DpBNVPC}, boxplots are computed over 5 runs of DP noise addition. The results show that the variations due to the randomness of the noise in our DP BN extractor are small.}
    \label{fig:DpBNVPCRandomNoiseEffect}
\end{minipage}
\end{figure*}
\section{Speaker versus Utterance-level X-Vector Selection}
\label{app:utterance_vs_speaker}

In our experiments, we use an utterance-level x-vector assignment strategy for all anonymization schemes, as well as for the design of our attacks (recall that attackers use the targeted anonymization scheme to anonymize the data they use to train their attack). Utterance-level assignment chooses a potentially different target x-vector for each utterance. Our specific utterance-level assignment strategy has the advantage of making the choice of x-vector independent of the speaker identity, thereby avoiding any leakage at this step. However, prior work also considered speaker-level assignment for anonymization (i.e., using the same target x-vector for all utterances of a given speaker) \cite{vpc2022csl,srivastava:hal-03197376}. This introduces a dependence on the speaker identity that may leak information and would need to be accounted for in a DP analysis.

In this section, we investigate the impact of such design choices empirically. Table~\ref{tab:PDRSvsPDRR} shows the empirical privacy for all possible combinations of assignment strategies in the anonymization scheme (here, we focus on \AnonPC) and the attack. We can draw two main conclusions: (i) speaker and utterance-level assignments provide the same level of protection against the best attack, and (ii) perhaps surprisingly, utterance-level assignment is the best choice for attacks, even when the anonymization scheme uses speaker-level assignment. This has led some prior work to largely overestimate the privacy protection provided by x-vector based speaker anonymization with speaker-level assignment, as recently observed in \cite{slicing}. We empirically found that the poor performance of speaker-level based attack is due to the ASI model overfitting the training data.

Overall, these results validate our choice of using utterance-level assignment for both the anonymization schemes and the attacks.

\section{Effect of the Randomness of DP Noise}
\label{sec:NoiseEffect}

Figure~\ref{fig:DpPitchVPCRandomNoiseEffectFinal} and Figure~\ref{fig:DpBNVPCRandomNoiseEffect} show the variations in privacy and utility due to the randomness of the noise in our DP extractors, rather than due to the randomness in the x-vector selection as in Figure~\ref{fig:DpPitchVPC} and Figure~\ref{fig:DpBNVPC}. We see that the variations due to the randomness of the noise are quite small (and typically smaller than those due to the randomness in x-vector selection).

\begin{table}[b] 
    \centering 
    \footnotesize
    \caption{Utterance-level DP guarantees using simple composition and advanced composition \cite{kairouz2015composition} for frame-level $\varepsilon=0.5$. KEY-- $K$: number of frames in an utterance.} 
       \begin{tabular}{lcccc} 
    \toprule  
    \multirow{2}{*}{\textbf{Composition}} & \multicolumn{4}{c}{\textbf{Utterance-level privacy budget} $\varepsilon$} \\ 
     & $K=100$ & $K=500$ & $K=1000$ & $K=10,000$ \\ 
    \hline
    \textbf{Simple} & 50 & 250 & 500 & 5,000\\
    \textbf{Advanced} with $\delta=10^{-5}$ & 36 & 114 & 198 & 1,464\\

    \bottomrule
    \end{tabular}
    \label{tab:compositions}
\end{table}
\section{Using advanced composition}
\label{app:tight}

We illustrate here how we can obtain tighter analytical privacy guarantees at the utterance-level by leveraging advanced composition theorems \cite{Dwork2014a,kairouz2015composition}. Specifically, 
Table~\ref{tab:compositions} reports the utterance-level DP guarantees for utterances with different length using the composition theorem of \cite[Theorem-3.4]{kairouz2015composition}. The results show that advanced composition can significantly improve the bound on the analytical privacy budget at the utterance level in comparison with using simple composition, especially for long utterances.